\documentclass[amsmath,amssymb,nofootinbib,prd]{revtex4}
\pdfoutput=1

\usepackage{graphicx}
\usepackage{epsfig}
\usepackage{rotate}
\usepackage{amsmath}
\usepackage{amssymb}
\usepackage{amsfonts}
\usepackage{afterpage}
\usepackage{color}
\usepackage{soul}

\def\hksqrt{\mathpalette\DHLhksqrt}
\def\DHLhksqrt#1#2{\setbox0=\hbox{$#1\sqrt{#2\,}$}\dimen0=\ht0
\advance\dimen0-0.2\ht0
\setbox2=\hbox{\vrule height\ht0 depth -\dimen0}%
{\box0\lower0.4pt\box2}}

\newcommand{\fnlt}{$f_{\rm NL}$ }
\newcommand{\txiss}{$\xi$ }
\newcommand{\tphi}{$\varphi$ }
\newcommand{\ttheta}{$\theta$ }

\begin{document}

\title{Large-scale 3D galaxy correlation function and non-Gaussianity}

\author{Alvise Raccanelli$^{1,2}$, Daniele Bertacca$^{3}$, Olivier Dor\'{e}$^{1,2}$, Roy Maartens$^{3,4}$\\}

\affiliation{
$^1$Jet Propulsion Laboratory, California Institute of Technology, Pasadena CA 91109, USA \\
$^2$California Institute of Technology, Pasadena CA 91125, USA \\
$^3$Physics Department, University of the Western Cape, Cape Town 7535, South Africa\\
$^4$Institute of Cosmology \& Gravitation, University of Portsmouth, Portsmouth PO1 3FX, UK}

\begin{abstract}

We investigate the properties of the 2-point galaxy correlation function at very large scales, including all geometric and local relativistic effects -- wide-angle effects, redshift space distortions, Doppler terms and Sachs-Wolfe type terms in the gravitational potentials. 
The general three-dimensional correlation function has a nonzero dipole and octupole, in addition to the even multipoles of the flat-sky limit.
We study how corrections due to primordial non-Gaussianity and General Relativity affect the multipolar expansion, and we show that they are of similar magnitude (when $f_{\rm NL}$ is small), so that a relativistic approach is needed.
Furthermore, we look at how large-scale corrections depend on the model for the growth rate in the context of modified gravity, and we  discuss how a modified growth can affect the non-Gaussian signal in the multipoles.

\end{abstract}

\date{\today}

\maketitle

\section{Introduction}

Some of the most important questions to be addressed by  forthcoming experiments in cosmology include testing models of the early Universe and understanding whether Einstein's General Relativity  is the correct theory for describing gravity. In both cases, the effects of possible deviations from the usually assumed (and observationally motivated) Gaussian initial conditions in the early Universe and the effects from relativistic corrections can be tested by measuring the clustering of matter on the largest scales observable. 
Future surveys will go deeper (e.g. the Prime Focus Spectrograph, PFS~\cite{pfs}) and wider (e.g. the Euclid satellite~\cite{Amendola:2012ys}), probing cosmological volumes around an order of magnitude larger than current ones and offering the possibility to observe the clustering of galaxies on scales comparable to the horizon scale, where both non-Gaussian effects and relativistic corrections become important.
This prospect requires a more accurate theoretical modeling on the largest scales. On these scales, we can no longer use the Newtonian and flat-sky approximations -- relativistic and geometric effects on large scales must be incorporated. 

The accurate modeling of large-scale matter clustering as traced by galaxies is the focus of this paper.
The clustering of galaxies as a means to test cosmological models and constrain parameters has a long history. Recent measurements are given in e.g.~\cite{Samushia:2011cs, samushiaboss, riedboss, sanchezboss, delatorrevipers, blakewigglez, contreraswigglez, beutler6df}; cosmological tests have been included in e.g.~\cite{Raccanelli:2012gt}.

In the standard flat-sky analysis~\cite{Kaiser:1987qv}, the angle $\theta$ between galaxies goes to zero (see Figure~\ref{fig:triangle}). This transverse flattening is also complemented by a flattening in the radial direction, i.e. the two galaxies are assumed to have the same average redshift.
Geometric corrections to the flat-sky approximation must break the transverse flattening, i.e. they must allow wide angles $\theta$ -- but they must also break the radial flattening, i.e. they must allow for large radial separation.
Wide-angle correlations have been investigated by \cite{Szalay:1997cc, Matsubara:1999du, Szapudi:2004gh, Papai:2008bd, Raccanelli:2010hk, Montanari:2012me}, and this has been applied to survey data by \cite{Samushia:2011cs, Raccanelli:2012gt}. 
However, for large-volume and deep surveys we are also interested in significant radial (redshift) separations. The full 3D correlation function, allowing for large separations in the transverse and radial directions, was analyzed in~\cite{Bertacca:2012tp}.

Relativistic effects arise from the fact that we observe galaxies on the past lightcone and not on a constant-time surface. The galaxy number overdensity is gauge-dependent and we have to compute the physical, observed overdensity. This introduces redshift, lensing and volume distortions, including the standard lensing and redshift distortions but with additional terms that can become significant on horizon scales.
The effects have been considered on the galaxy power spectrum~\cite{Yoo:2009au, Yoo:2010ni, Bonvin:2011bg, Challinor:2011bk} and on the 2-point correlation function~\cite{Bertacca:2012tp}. The combination of relativistic effects and primordial non-Gaussianity has also been investigated~\cite{Bruni:2011ta, Jeong:2011as, Maartens:2012rh, Yoo:2012se}.
Horizon scales are also where dark energy and modified gravity have the strongest impact on clustering, hence it is important to incorporate the relativistic effects for tests of modified gravity and dynamical dark energy~\cite{Hall:2012wd, Lombriser:2013aj, Duniya:2013eta}. 

The work presented here is based on~\cite{Bertacca:2012tp}, which introduced a fully relativistic and 3D formalism that recovers and generalizes previous work in the Newtonian flat-sky~\cite{Kaiser:1987qv, Hamilton:1997zq}, Newtonian wide-angle~\cite{Szalay:1997cc, Matsubara:1999du, Szapudi:2004gh, Papai:2008bd, Raccanelli:2010hk, Samushia:2011cs, Raccanelli:2012gt, Montanari:2012me} and relativistic flat-sky \cite{Yoo:2009au, Jeong:2011as, Yoo:2010ni} cases.
We refer to the correlation function as the ``3D correlation function''. The operator that includes all the terms of the Jacobian has been called the ``radial operator''~\cite{Hamilton:1997zq} or the ``spherical operator''~\cite{Hamilton:1995px, Bharadwaj:1998bq} .

We point out here that the flat-sky limit suppresses the odd multipoles -- a dipole and octupole -- of the correlation function, and that these multipoles are potentially important for probing large-scale clustering. The full geometric effects should be combined with the relativistic effects for a consistent and accurate analysis. Since non-Gaussianity also grows on large scales, it is important to take relativistic and geometric effects into account, as we do here.  

The paper is organized as follows. In Section~\ref{sec:xi} we briefly describe the formalism we use for modeling the large-scale correlation function while in Section~\ref{sec:nGxi} we discuss the modifications to the correlation function due to primordial non-Gaussianity. In Section~\ref{sec:multipoles} we show that for the 3D case, additional odd multipoles are present in the correlation function compared to the flat-sky approximation. In Section~\ref{sec:nG_relativistic } we show how non-Gaussianity is imprinted in the multipoles, and how this can be used to distinguish between non-Gaussianity and relativistic effects. In Section~\ref{sec:growth} we illustrate how the non-Gaussian correlation function is affected by modified gravity. Finally we discuss our conclusions in Section~\ref{sec:conclusions}.

\section{3D Correlation Function: including large angular and radial separations}
\label{sec:xi}

\begin{figure}[!htbp] 
\vspace*{-1.5cm}
\includegraphics[width=0.5\linewidth]{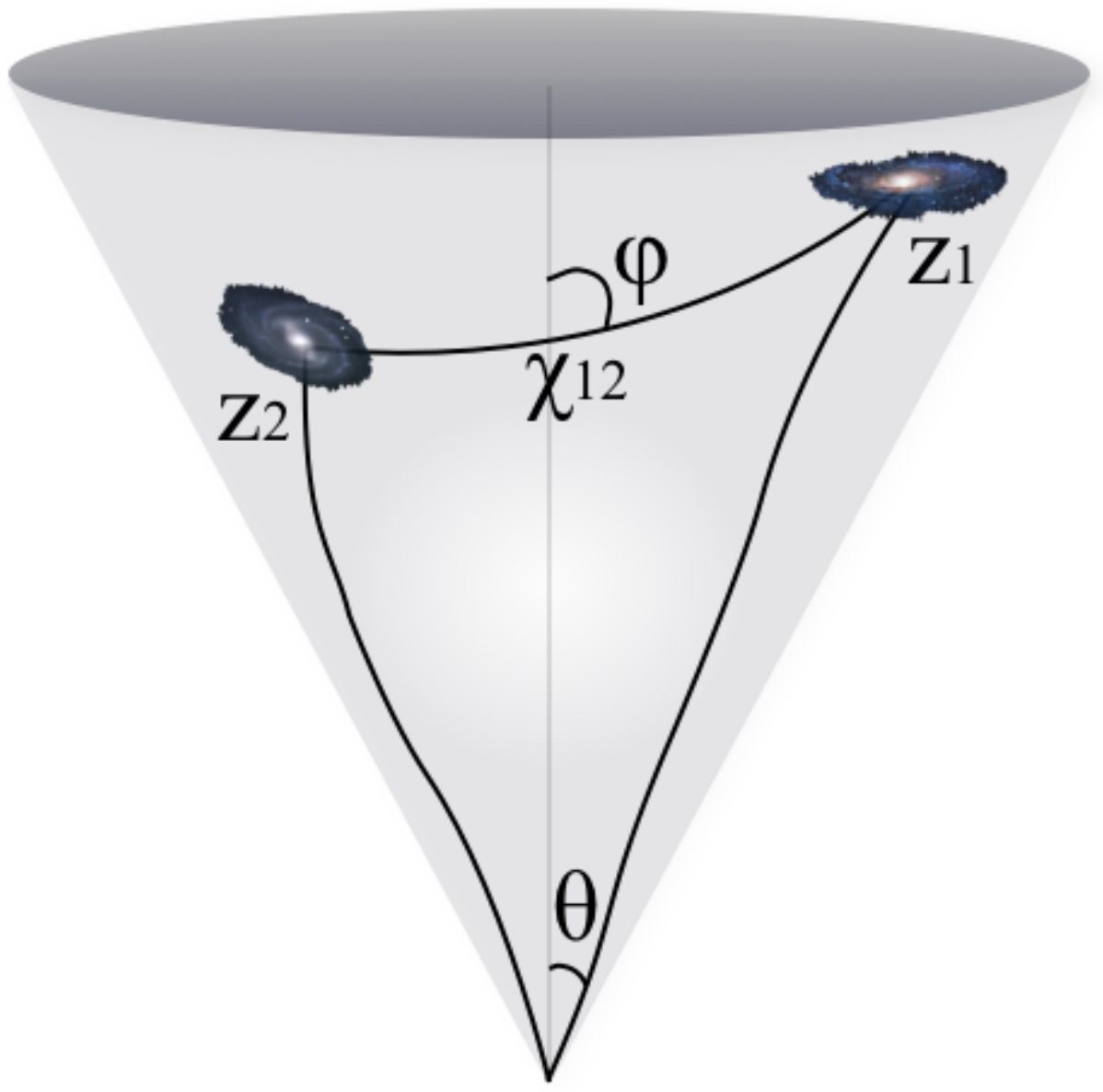}
\vspace*{-1.5cm}
\caption{Geometry of the problem: the triangle formed by the observer and the pair of galaxies.}
\label{fig:triangle}
\end{figure}

In~\cite{Bertacca:2012tp} a formalism is developed for computing the 2-point correlation function, including all effects due to the particular geometry of the problem illustrated in Figure~\ref{fig:triangle}, as well as all relativistic effects arising from observations on the past lightcone.

When computing the correlation function on scales comparable to the horizon, we need to deal with gauge dependence of perturbed quantities (which is not a problem on sub-horizon scales). The solution is to identify the observed galaxy number density contrast $\Delta_{\rm obs}$ (which is necessarily gauge-invariant)~\cite{Yoo:2009au, Yoo:2010ni, Bonvin:2011bg, Challinor:2011bk}. Then $\Delta_{\rm obs}$ can be computed in any gauge. 

In general, it has the form:
\begin{equation}
\label{eq:Delta}
\Delta_{\rm obs} = \Delta_{\rm loc}+\Delta_{\rm int} \, ,
\end{equation}
where  $\Delta_{\rm loc}$ includes the galaxy density contrast and all other local terms evaluated at the source, and $\Delta_{\rm int}$ includes the lensing convergence and all other terms that are line of sight integrals (for details, see~\cite{Bertacca:2012tp}).
In longitudinal gauge, $\Delta_{\rm loc}$
contains the Kaiser redshift distortion, additional velocity terms and Sachs-Wolfe type terms in the gravitational potentials, while $\Delta_{\rm int}$ contains the lensing convergence and time delay integrals along the line of sight. 

Here we focus on the local term. A further analysis including the integrated terms is left for future work \cite{Raccanelliradial}.  
The auto-correlation function of the local part of the density contrast in redshift space is
\begin{eqnarray}
\label{eq:xiss}
\xi_{\rm loc}( {\bf x}_1, z_1, {\bf x}_2, z_2) = \langle \Delta_{\rm loc}( {\bf x}_1, z_1) \Delta_{\rm loc}({\bf x}_2, z_2) \rangle \, ,
\end{eqnarray}
where ${\bf x}= {\bf n}\,\chi(z)$, with  $\chi$ the comoving distance and  ${\bf n}$  the unit vector in the direction of the galaxy. The observed galaxy number density contrast is \cite{Bertacca:2012tp}
\begin{eqnarray}
\label{eq:delta_s}
\Delta_{\rm obs} ({\bf x}, z) = b(z) \left\{ \left[1+\frac{1}{3}\beta(z) \right]
\mathcal{A}^0_0({\bf x}, z) +\frac{2}{3}\beta(z) \mathcal{A}^0_2({\bf x}, z) + \frac{\alpha(z)\beta(z)}{\chi(z)}
\mathcal{A}^1_1({\bf x}, z) + \gamma(z) \mathcal{A}^2_0({\bf x}, z) \right\} \, ,
 \end{eqnarray}
where $\mathcal{A}^n_\ell$ are spherical transforms of $\delta$~\cite{Szalay:1997cc}:
\begin{equation}
\label{A-tensor} \mathcal{A}^n_\ell ({\bf x}, z)= \int
\frac{d^3k}{(2\pi)^3} (ik)^{-n}  \, \mathcal{P}_\ell \left({{\bf n} \cdot {\bf k} \over k}\right)\exp{\left(i{\bf k} \cdot {\bf x}\right)}\:
\delta({\bf k},z) \, .
\end{equation}
Here $\mathcal{P}_\ell$ are Legendre polynomials and $\delta$ is the matter overdensity in synchronous-comoving gauge (which is gauge invariant). The galaxy bias $b$ is defined in this gauge (for a careful discussion on this issue, see e.g.~\cite{Bruni:2011ta, Jeong:2011as}). 

The terms $\alpha$, $\beta$ and $\gamma$ have the following meaning (see~\cite{Bertacca:2012tp} and also \cite{Raccanelli:2010hk} for details): 
\begin{itemize}
\item
 $\alpha$ encodes the ``mode-coupling'' effect, which mixes different modes~\cite{Raccanelli:2010hk} of the wide-angle correlation, and also includes volume distortion on the lightcone. The third term in Equation~\eqref{eq:delta_s} thus describes geometry and also relativistic corrections.
\item
 $\beta$ encodes the effect of the matter overdensity $\delta$ and peculiar velocity. The first two terms ($\beta$ terms) in Equation~\eqref{eq:delta_s} appear in the usual Newtonian flat-sky approximation.
\item
 $\gamma$ encodes the effect of the gravitational potentials.  The last term in Equation~\eqref{eq:delta_s} is a purely relativistic term that is not present in the Newtonian treatment.
\end{itemize}

The explicit forms are: 
\begin{eqnarray}
\label{eq:alpha}
\alpha(z)&=& \alpha_{\rm Nwt} (z) -  \frac{\chi(z) H(z)}{(1+z)} \left[\frac{3}{2}\Omega_m (z) -
1-2\mathcal{Q}(z)\right] \; , \\
\label{eq:beta}
\beta(z)  &=& \frac{f(z)}{b(z)},  ~~~~ f=-{d \ln D(z) \over d \ln (1+z)}\;, \\
\gamma(z) &=& \frac{H(z)}{(1+z)} \left\{\frac{H(z)}{(1+z)}
\left[\beta(z) -\frac{3}{2}\frac{\Omega_m (z)}{b(z)}
\right]b_e(z) + \frac{3}{2}\frac{H(z)}{(1+z)} \beta(z) \big[
\Omega_m (z) - 2 \big]
\right.  \nonumber \\
& & \left. -\frac{3}{2} \frac{H(z)}{(1+z)}  \frac{\Omega_m
(z)}{b(z)} \left[1-4\mathcal{Q}(z) + \frac{3}{2}\Omega_m
(z)\right] +\frac{3}{\chi(z)}
\big[1-\mathcal{Q}(z)\big]\frac{\Omega_m (z)}{b(z)} \right\} ,
\label{eq:gammaz}
\end{eqnarray}
where $D(z)$ is the linear growth factor and $\alpha_{\rm Nwt}$ is the Newtonian part of $\alpha$:
\begin{equation}
\label{eq:alpha_nwt}
{\alpha_{\rm Nwt}(z) \over  \chi(z)}= - \frac{H(z)}{(1+z)}\left\{b_e(z)-\frac{2}{\chi(z)}\big[1-\mathcal{Q}(z)\big]\frac{(1+z)}{H(z)}\right\}=  \frac{d \ln{N_g}}{d \chi}+ \big[1-\mathcal{Q}(z)\big]\frac{2}{\chi}\;.
\end{equation}

The magnification bias and evolution bias are~\cite{Jeong:2011as}:
\begin{equation}
\mathcal{Q} = - {\partial \ln N_g \over \partial \ln \mathcal{L}}, ~~~ b_e =-(1+ {z}) {\partial\ln N_g \over \partial z},
\end{equation}
where $N_g$ is the comoving number density of galaxies of luminosity $\mathcal{L}$.
In this work we consider a Euclid-like survey~\cite{Amendola:2012ys} for $b(z)$ and $N_g(z)$.
For simplicity, the magnification bias $\mathcal{Q}$ is set to zero. Different surveys will have different values of $\mathcal{Q}$ (possibly varying with redshift).

In the Newtonian case, the observed number density contrast becomes:
\begin{eqnarray}
\label{eq:delta_nwt}
{\Delta_{\rm Nwt}  ({\bf x}, z)} = \left[b(z) + \frac{1}{3}f(z) \right] \mathcal{A}^0_0({\bf x}, z) 
+ \frac{2}{3}f(z) \mathcal{A}^0_2({\bf x}, z)
+ \frac{f(z)\alpha_{\rm Nwt} (z)}{\chi(z)} \mathcal{A}^1_1({\bf x}, z) \,.
 \end{eqnarray}
For  $\mathcal{Q}=0$ we recover the form in \cite{Hamilton:1997zq} and if $\mathcal{Q}= 0$ and $N_g$ is constant, then $\alpha_{\rm Nwt} =2$, as in \cite{Szapudi:2004gh, Papai:2008bd}.

Following the formalism of~\cite{Szalay:1997cc, Szapudi:2004gh, Papai:2008bd, Raccanelli:2010hk, Bertacca:2012tp}, we decompose the correlation function as the finite sum:
\begin{equation}
\label{eq:xi_ss}
{\xi}( {\bf x}_1, z_1, {\bf x}_2, z_2) = b(z_1) b(z_2)
\sum_{\ell_1,\ell_2,L,n}
B_{n}^{\, \ell_1\ell_2L}( {\chi}_1, {\chi}_2)\, 
S_{\ell_1\ell_2L}( {{{\bf n}}}_1, { {{\bf n}}}_2,  { {{\bf n}}}_{12}) \,
\xi_L^{\, n}(\chi_{12}; z_1, z_2) \, ,
\end{equation}
where ${\bf x}_{12}= {\bf x}_1- {\bf x}_2 \equiv \chi_{12} {\bf n}_{12}$, and $S_{\ell_1\ell_2L}$ are tripolar spherical harmonics \cite{Szalay:1997cc, Szapudi:2004gh, Papai:2008bd}.  The $B_{n}^{\, \ell_1\ell_2L}$ coefficients contain the corrections due to the functions $\alpha$, $\beta$ and $\gamma$~\cite{Bertacca:2012tp}. The functions:
\begin{equation}
\label{eq:xiLn}
\xi_L^{\, n}(\chi; z_1, z_2) = \int \frac{k^{2-n}}{2\pi^2}
\, j_L(\chi k) \, P_\delta(k; z_1, z_2) \, dk \; ,
\end{equation}
are spherical Bessel transforms of the matter power spectrum $P_\delta(k; z_1, z_2)$~\cite{Szalay:1997cc}.

\section{Primordial non-Gaussianity and the correlation function}
\label{sec:nGxi}

Deviations from Gaussian initial conditions in the local (squeezed) limit are usually parametrized by the dimensionless parameter $f_{\rm NL}$ in the Taylor expansion:
\begin{equation}
\label{eq:fnl}
\Phi=\phi+f_{\rm NL}\left(\phi^2-\langle\phi^2\rangle\right) ,
\end{equation}
where $\Phi$ is the gauge-invariant gravitational potential and $\phi$ is a Gaussian random field.
Nonzero $f_{\rm NL}$ introduces a scale-dependent modification of the large-scale halo bias (see e.g. \cite{Matarrese:2000iz, Dalal:2007cu, Matarrese2008,Slosar:2008hx, Desjacques:2010jw, Xia:2010pe}): 
\begin{equation}
\label{eq:ng-bias}
b(z)~ \to ~ b(z)+\Delta b(z, k) =b(z)+ [b(z)-1] f_{\rm NL}\delta_{\rm ec} \frac{3 \Omega_{m0}H_0^2}{c^2k^2T(k)D(z)}, 
\end{equation}
where $T(k)$ is the matter transfer function ($\to 1$ on large scales) and $\delta_{\rm ec}$ is the critical value of the matter overdensity for ellipsoidal collapse ($\delta_{\rm ec}=\delta_{\rm c}\hksqrt{q}$, with $q$ a parameter fitted via simulations~\cite{Xia:2010pe}).
Here $b(z)$ is the usual bias calculated assuming Gaussian initial conditions. We assume it to be scale-independent, as we concentrate on large scales (for discussion of the issue of large-scale bias, see~\cite{Matarrese97, Moscardini98, Hamaus10, Baldauf11, Baldauf13}).

Local non-Gaussianity affects the correlation function of Equation~\eqref{eq:xi_ss} via the bias factors: $b (z_i)\to  b(z_i)+\Delta b(z_i,k)$. The nonlinear parameter $f_{\rm NL} $ in the potential thus leads to a significant correction to the linear galaxy power on large scales via the bias -- as a result of a long-wavelength modulation of the short-wavelength modes responsible for halo collapse.

Most measurements of non-Gaussianity from clustering analyses so far rely on the above scale-dependent bias and have been performed using the power spectrum of either radio or optical galaxies in a flat-sky and Newtonian approximation (see e.g. \cite{Slosar:2008hx, Xia:2010pe, Xia:2011hj, Ross:2012sx}). 

Following~\cite{Bruni:2011ta, Bertacca:2012tp}, here we will use a fully 3D and relativistic approach to revisit the correlation function computation in the non-Gaussian case.
Recently, the Planck satellite~\cite{Ade:2013ydc} found $f_{\rm NL}=2.7 \pm 5.8$, thus constraining strongly the amount of primordial non-Gaussianity. This constraint is unlikely to be significantly improved upon by future CMB experiments, and so it is important to pursue large-scale galaxy surveys (or 21cm intensity mapping surveys~\cite{Camera:2013kpa}) as the means to improve the constraint. Since the local non-Gaussianity is small, it is comparable to the relativistic effects in the observed clustering~\cite{Bruni:2011ta}, and thus these effects, together with geometric effects, should be incorporated in analysis of future survey data. (Note that even if the primordial fluctuations are exactly Gaussian, there will be a nonzero effective local non-Gaussianity on large scales from relativistic nonlinearity in the curvature perturbation, giving $f_{\rm NL}=-5/3$~\cite{Verde:2009hy}.)

\section{Odd Multipoles beyond the flat-sky limit}
\label{sec:multipoles}

The correlation function is usually expanded in multipoles using Legendre polynomials. In the flat-sky approximation, this results in a sum over the first three even multipoles \cite{Hamilton:1997zq}, the other multipoles vanishing.
However, in the 3D case, where the redshift distortion operator destroys the translational symmetry, the odd multipoles are no longer zero, and this enables us to extract additional angular information from a redshift distortion analysis. 

\begin{figure}
\includegraphics[width=0.23\linewidth]{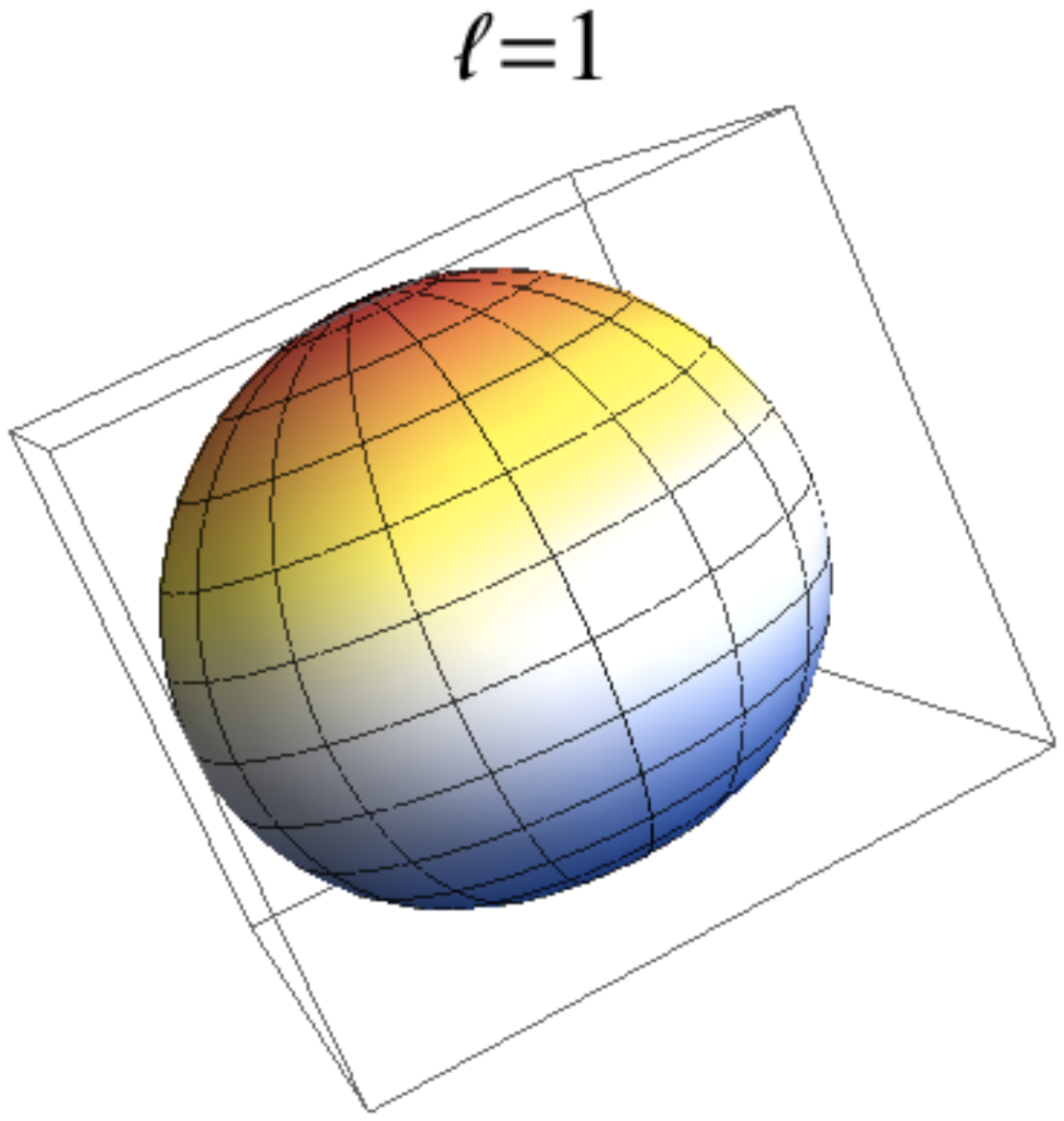}
\includegraphics[width=0.23\linewidth]{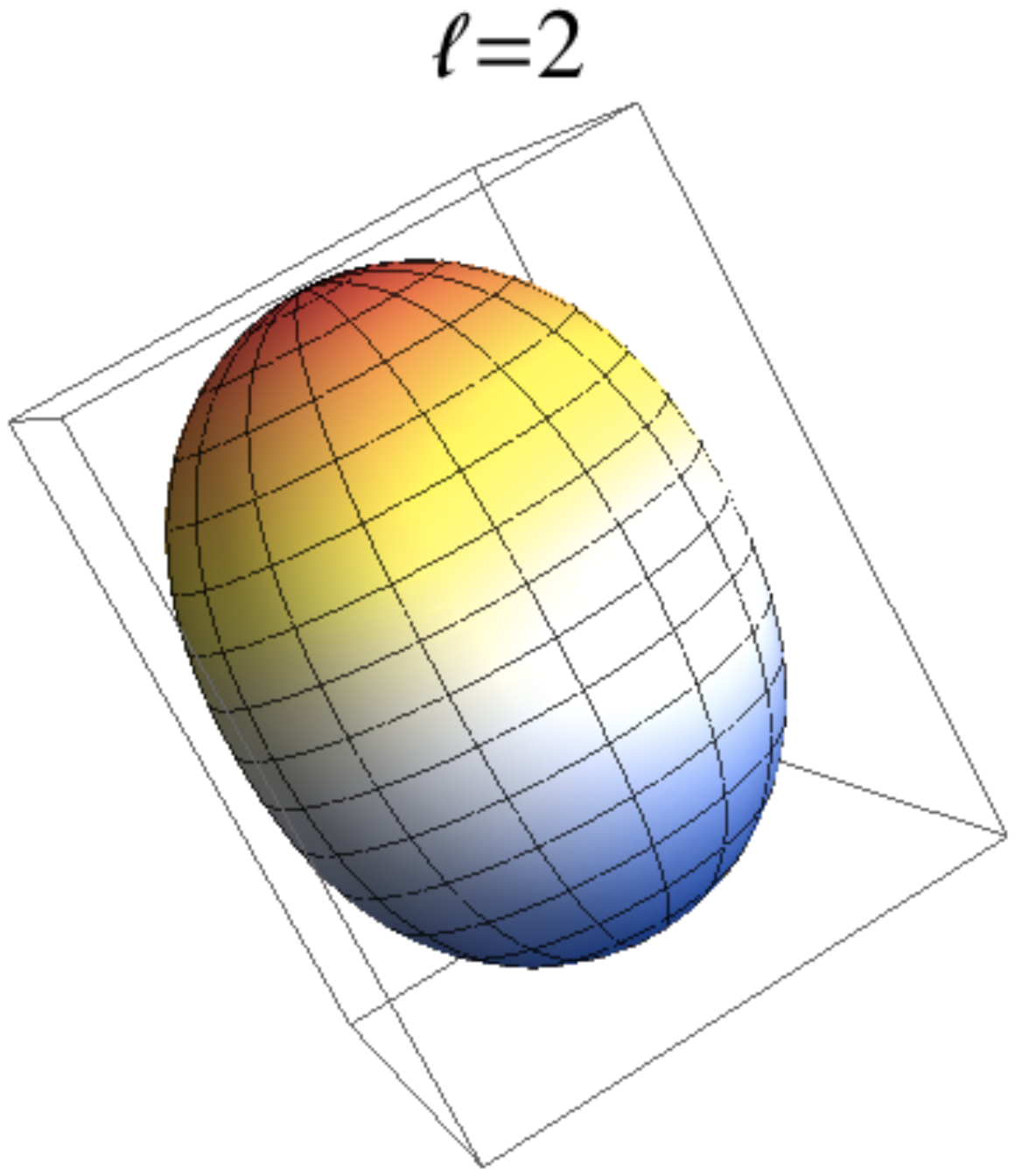}
\includegraphics[width=0.23\linewidth]{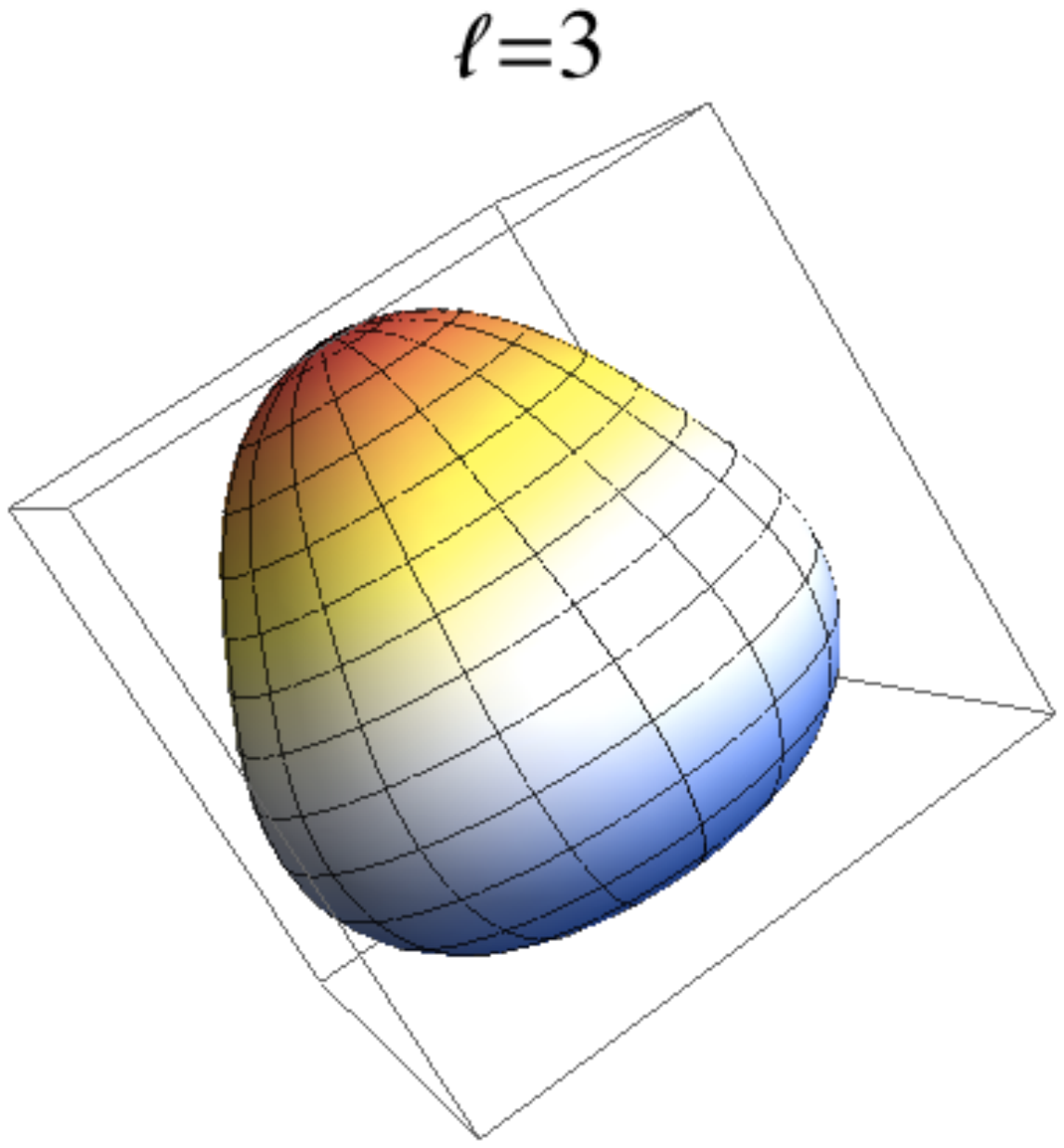}
\includegraphics[width=0.23\linewidth]{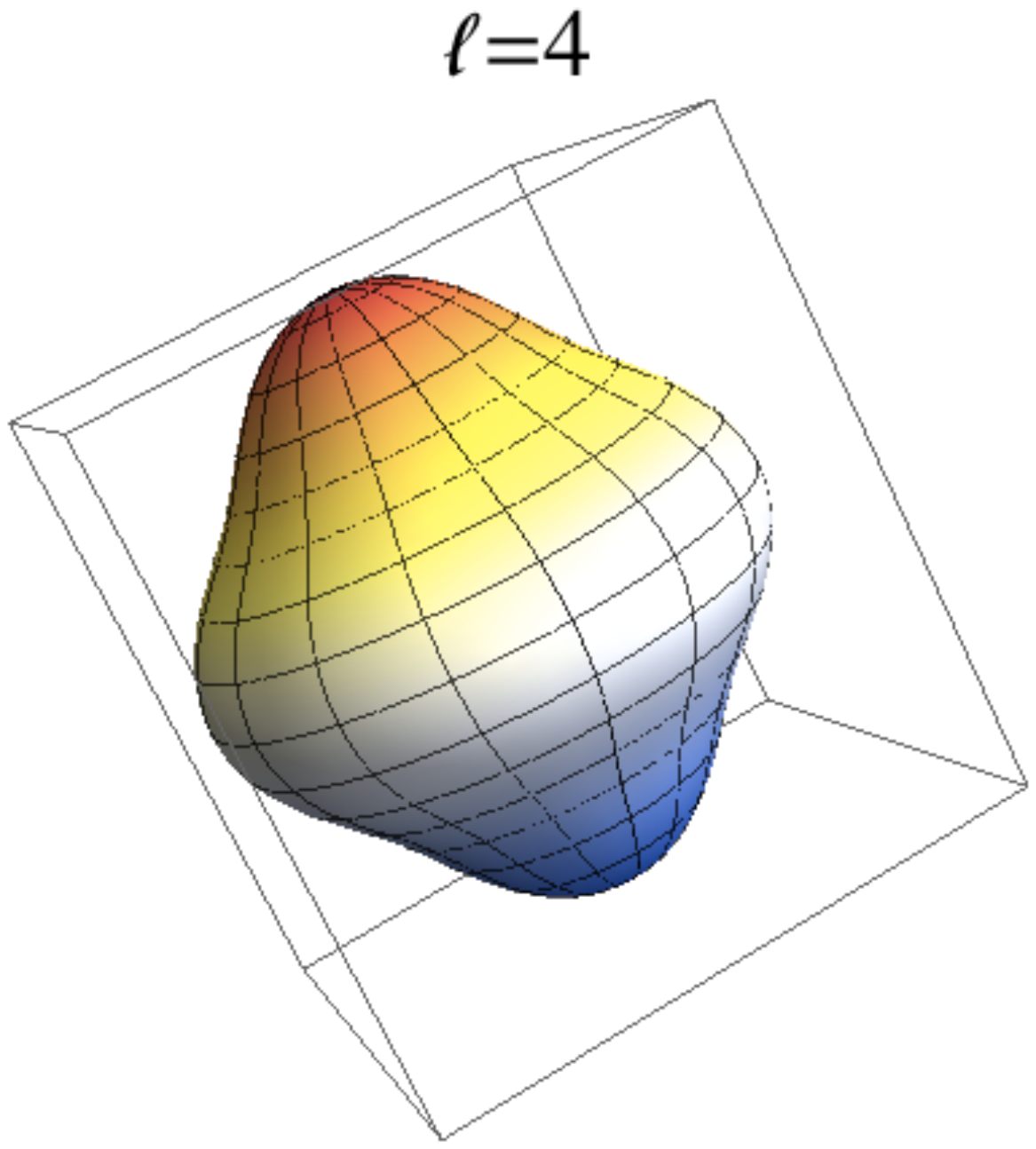}
\caption{Shapes of the multipoles as perturbations of a sphere. From left to right: dipole ($L = 1$), quadrupole ($L= 2$), octupole ($L = 3$) and hexadecapole ($L = 4$). The monopole only modifies the radius and is not shown.}
\label{fig:xi_ells}
\end{figure}

\begin{figure}
\includegraphics[width=0.47\linewidth]{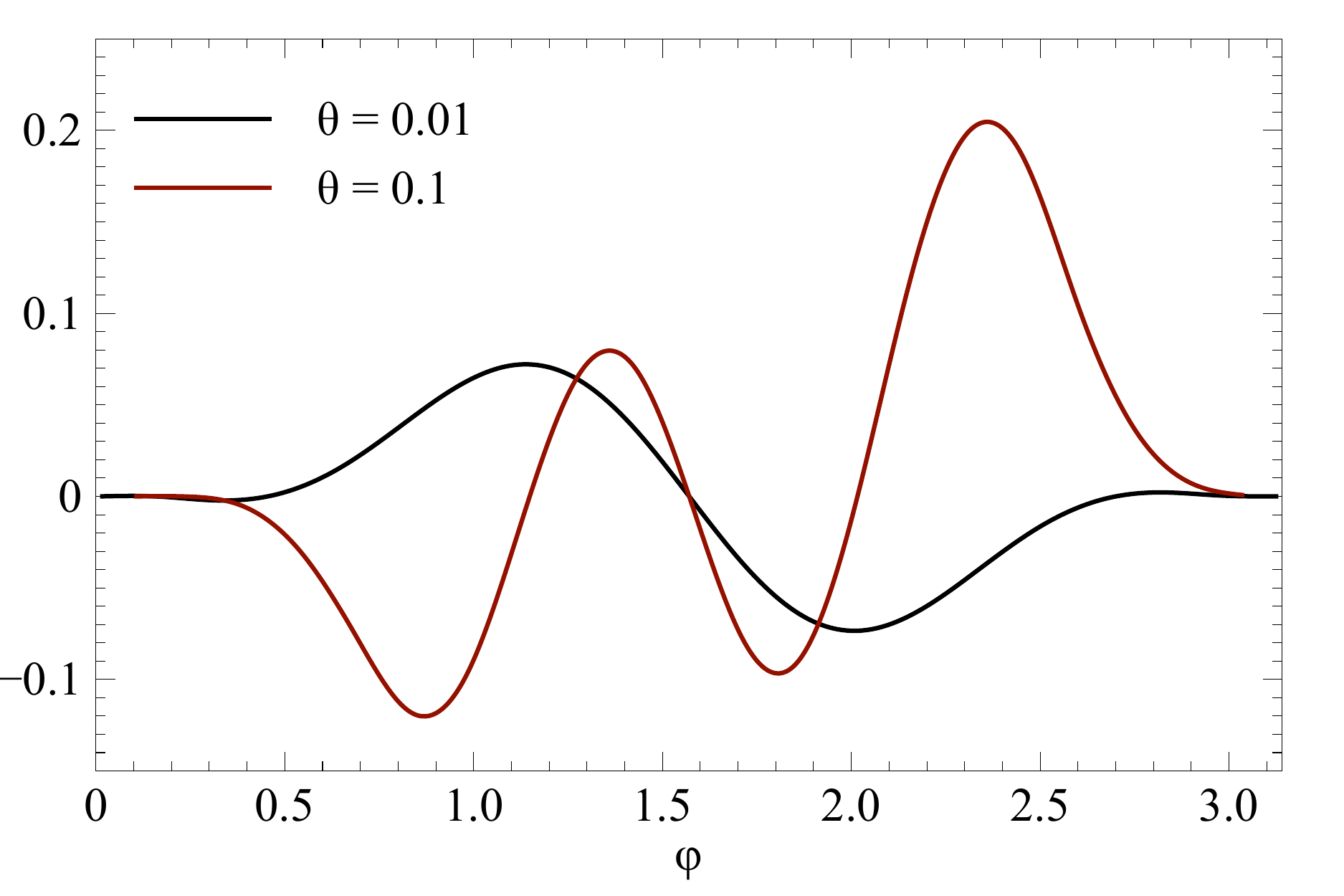}
\includegraphics[width=0.47\linewidth]{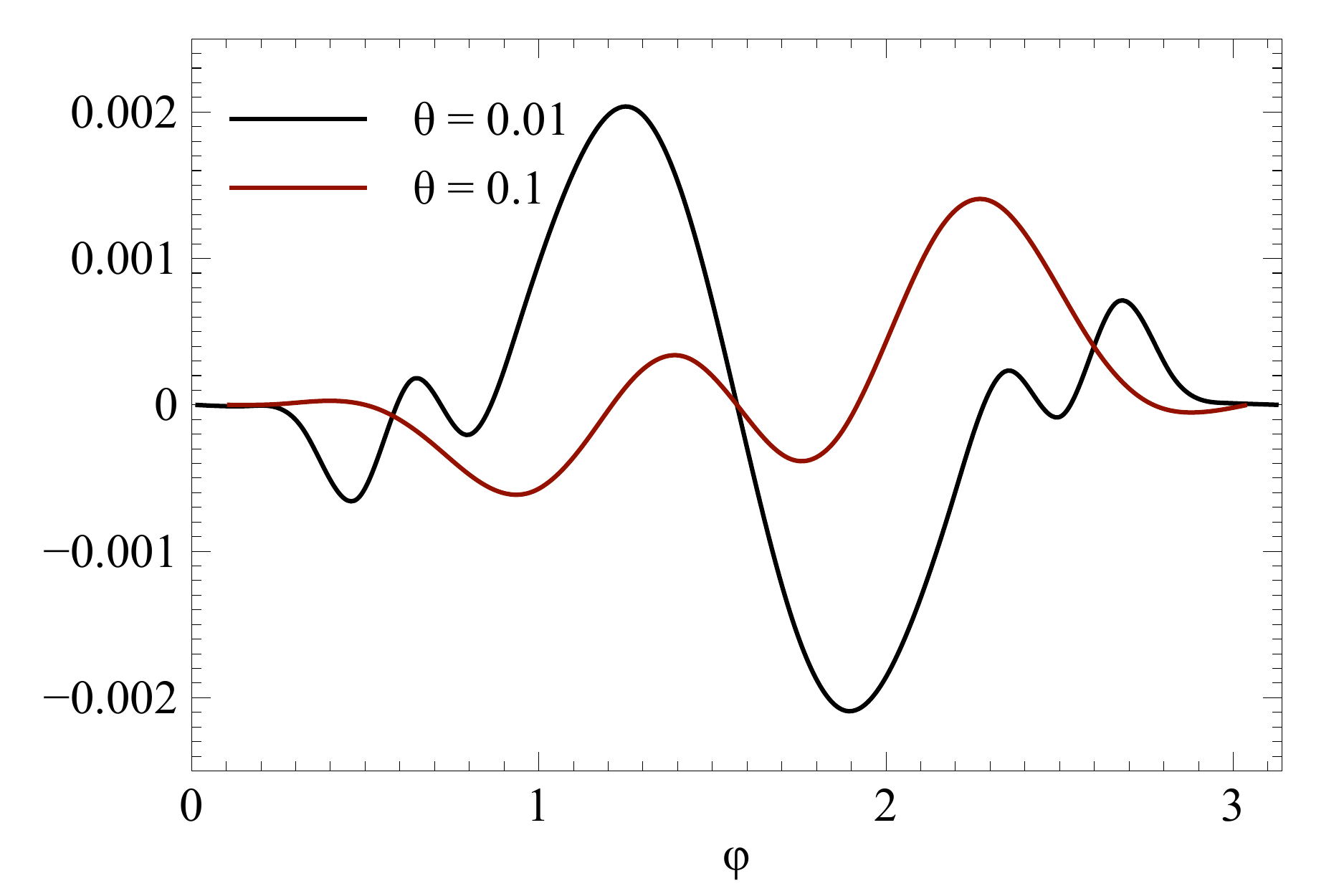}
\caption{The integrand of Equation~\eqref{eq:multip_standard} as a function of the pair orientation angle, in the dipole case ($L=1$), for $\theta=0.01, 0.1$, and for a fixed redshift $z=0.3$ ({\it left}) and $z=1.5$ ({\it right}). The integrand for the larger $\theta$ has been amplified by a constant factor for illustrative purposes (because different $\theta$ correspond to different linear separation scales). Note the loss of symmetry with respect to $\varphi=\pi/2$ when $\theta \neq 0$.}
\label{fig:xi_1}
\end{figure}

\begin{figure}
\includegraphics[width=0.47\linewidth]{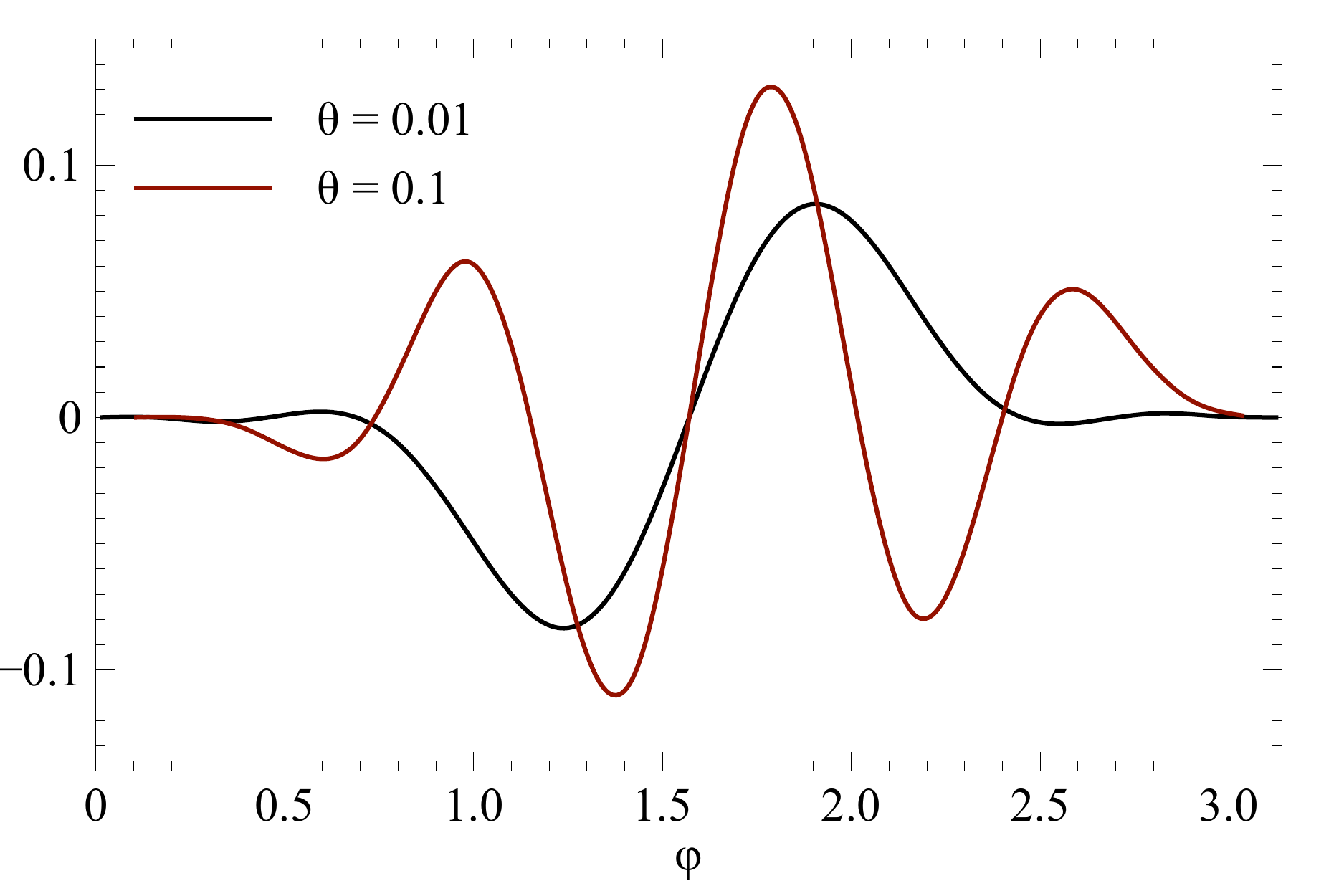}
\includegraphics[width=0.47\linewidth]{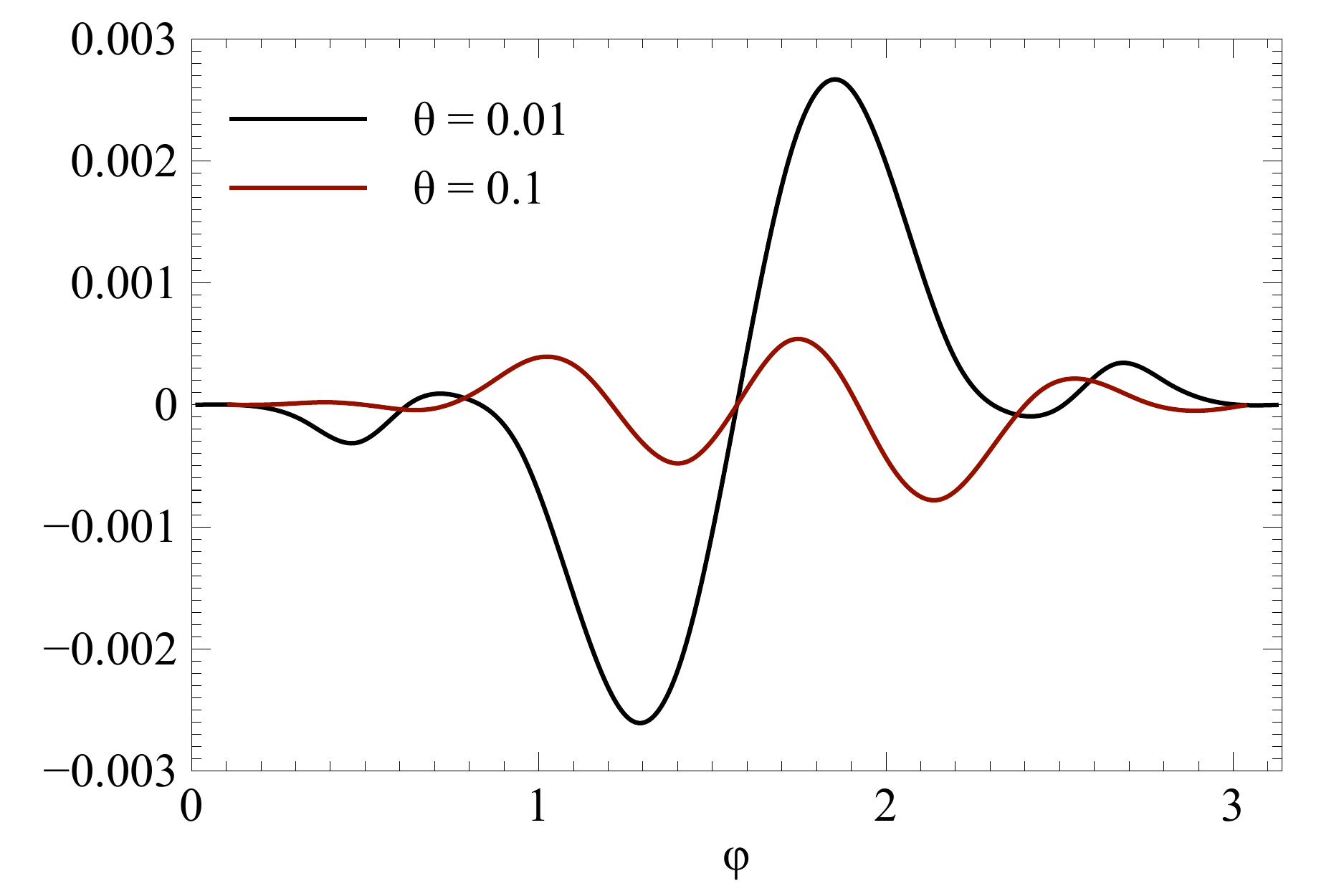}
\caption{As in Fig. \ref{fig:xi_1}, for the octupole case $L=3$.}
\label{fig:xi_3}
\end{figure}

The 3D correlation function is a function of 3 variables, and it is natural to write it as  $\xi(z_1, z_2, \theta)$~\cite{Szalay:1997cc, Papai:2008bd, Bertacca:2012tp}. Here we are explicitly considering integrations over $\varphi$, so we write the correlation function as $\xi(z_2, \theta, \varphi)$. Given a set of $\{z_2, \theta, \varphi \}$, $z_1$ is automatically defined by the geometry of the system via (see Figure~\ref{fig:triangle}):
\begin{eqnarray}
\varphi(z_1,z_2,\theta) &=& \cos^{-1}\left\{ \sqrt{\frac{1+\cos 2 \theta}{2}}\left[\frac{\chi(z_1)-\chi(z_2)}{\chi_{12}(z_1,z_2,\theta)}\right]\right\} \, ,\\
\chi_{12}(z_1,z_2,\theta)&=&\sqrt{\chi(z_1)^2-2 \chi(z_1)\chi(z_2)\cos 2\theta+\chi(z_2)^2}\,.
\end{eqnarray}
The flat-sky approximation loses the real 3D structure of the system, whereas in our 3D case we can capture its geometry, and this allows a more precise investigation of the structure of the density field. In Figure~\ref{fig:xi_ells} we illustrate the shape of the different multipoles by showing the deformations on a sphere induced by them. While the quadrupole and hexadecapole have previously been shown (e.g. Figure 6 of~\cite{Hamilton:1997zq}), the odd multipoles are zero in the flat-sky limit and are usually neglected.

The expansion in Legendre polynomials is obtained by integrating the correlation function over the pair orientation angle $\varphi$:
\begin{equation}
\label{eq:multip_standard}
\xi_{L}(z_2, \theta) = \frac{2 L +1}{2} \frac{\pi}{\pi-2\theta} \int_{\theta}^{\pi-\theta} ~ {\rm d} \, \varphi ~ \xi(z_2, \theta, \varphi) ~ \mathcal{P}_L 
 \left\{\cos\left[\frac{\pi \left( \varphi - \theta \right)}{\pi - 2\theta}\right]\right\} \, \rm sin \left[\frac{\pi \left( \varphi - \theta \right)}{\pi - 2\theta}\right]  .
\end{equation}
To ensure orthogonality of Legendre polynomials in the case of $\theta \neq 0$, we have introduced suitable modifications in the angular dependence of the multipole expansions -- i.e. the argument of the Legendre polynomials must be modified as above. 
This arises because the integration range is different from that in the flat-sky limit, due to the geometric constraints that appear in the general case, and because the radial part of the operator modifies the properties of the system.
Note that
\begin{eqnarray}
\label{ort-legendreNEW}
\frac{2 L +1}{2} \frac{\pi}{\pi-2\theta} \int_{\theta}^{\pi-\theta} ~ {\rm d} \, \varphi ~  \mathcal{P}_{\tilde{L}}(\cos\varphi)  \mathcal{P}_L \left\{\cos\left[\frac{\pi \left( \varphi - \theta \right)}{\pi - 2\theta}\right]\right\}\, \sin \left[\frac{\pi \left( \varphi - \theta \right)}{\pi - 2\theta}\right]\,  =
\left\{ \begin{array}{ll}
\neq 0 & {\rm for}~\tilde{L}+L {\rm \;even} \\ \\
\simeq 0 & {\rm for}~\tilde{L}+L {\rm \;odd}
\end{array} \right.  
\end{eqnarray}
In Appendix A we discuss alternative attempts at finding the correct multipole expansion.

The odd multipoles do not integrate to 0 in the 3D case since symmetry about $\varphi=\pi/2$ is lost compared to the flat-sky case, where the two redshifts of the galaxies are assumed to be the same. This is shown explicitly in Figures~\ref{fig:xi_1}-\ref{fig:xi_3}, where we plot the integrand of Equation~\eqref{eq:multip_standard} for the first and third multipoles, with a small ($=0.01$) and a relatively large ($=0.1$) value of $\theta$, as a function of the pair orientation angle $\varphi$. As we move away from small $\theta$, the symmetry of the integrand around $\varphi=\pi/2$ is destroyed.

The 3D case is different from the flat-sky limit also in the mathematical description of the problem -- having two angles makes the usual expansion in Legendre multipoles slightly different. 
With the expansion in tripolar spherical harmonics as suggested by~\cite{Szalay:1997cc}, we can express the correlation function using different formalisms (see also~\cite{Szapudi:2004gh, Papai:2008bd, Raccanelli:2010hk}).
One involves the expansion of Equation~\eqref{eq:xi_ss}, and in this case the coefficients of the tripolar spherical harmonics $B_{n}^{\, \ell_1\ell_2L}$ have been derived and presented in~\cite{Bertacca:2012tp}, in the relativistic case (for two alternatives and their coefficients, in the Newtonian case, see~\cite{Papai:2008bd}).

\begin{figure}
\includegraphics[width=0.47\linewidth]{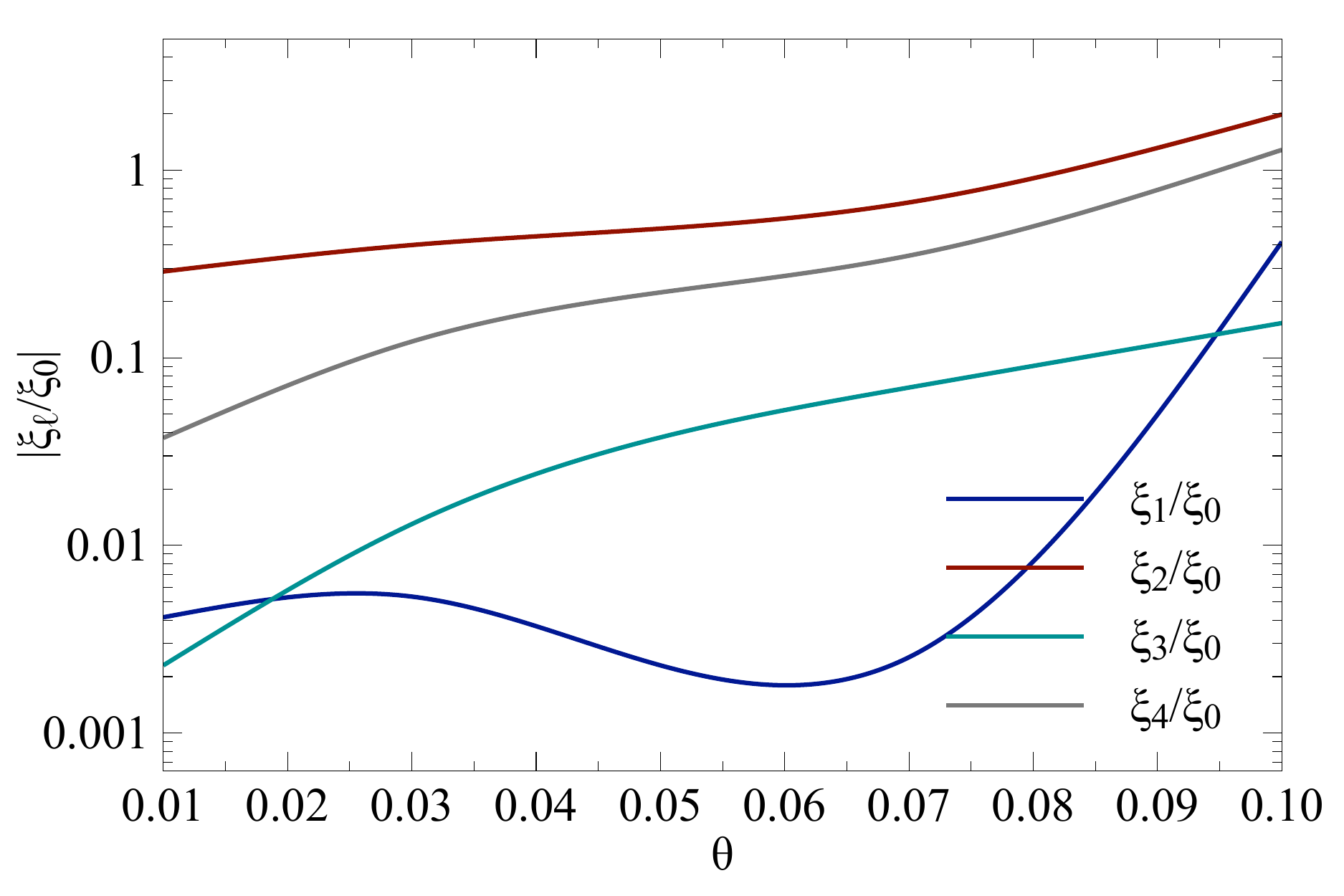}
\includegraphics[width=0.47\linewidth]{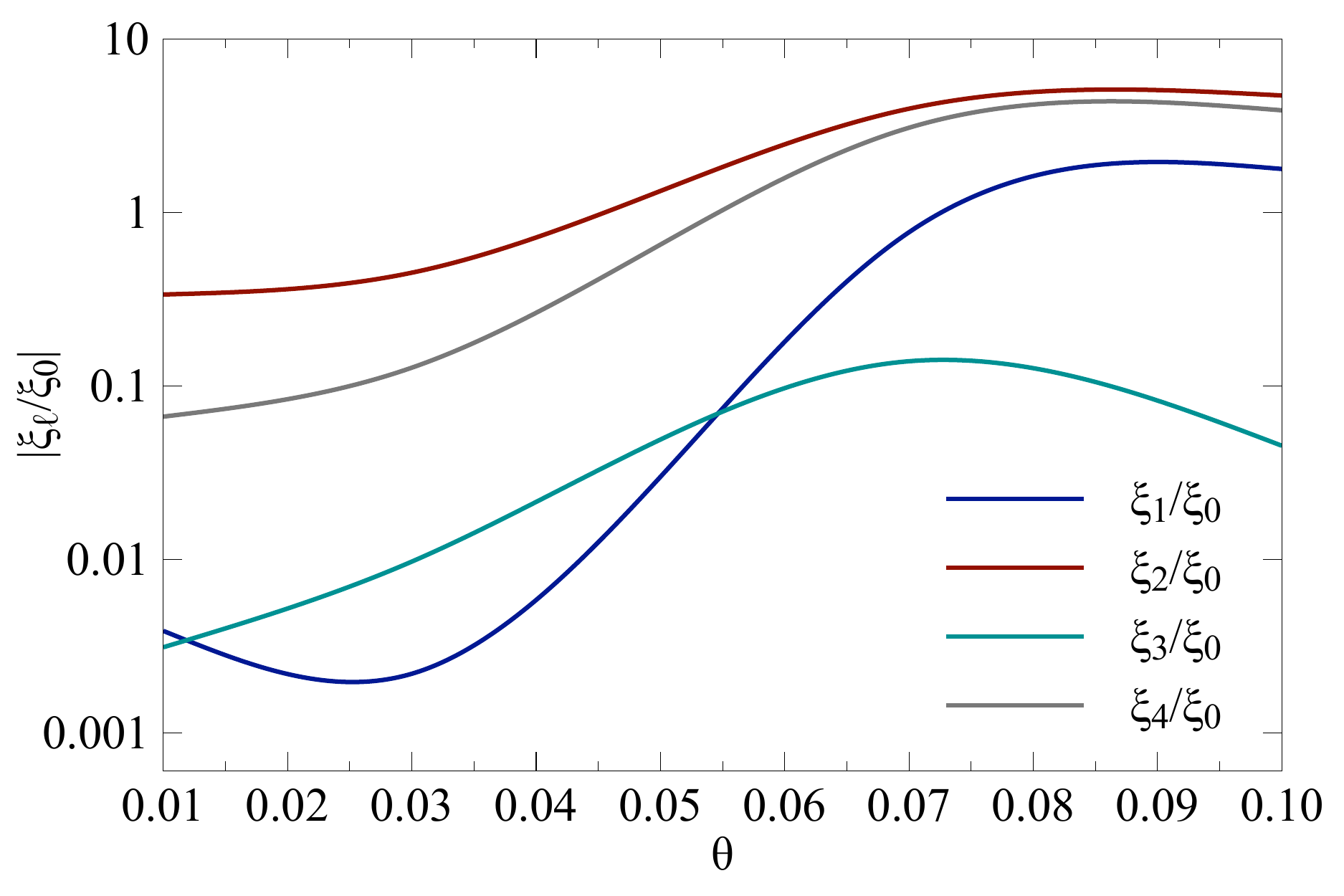}
\includegraphics[width=0.47\linewidth]{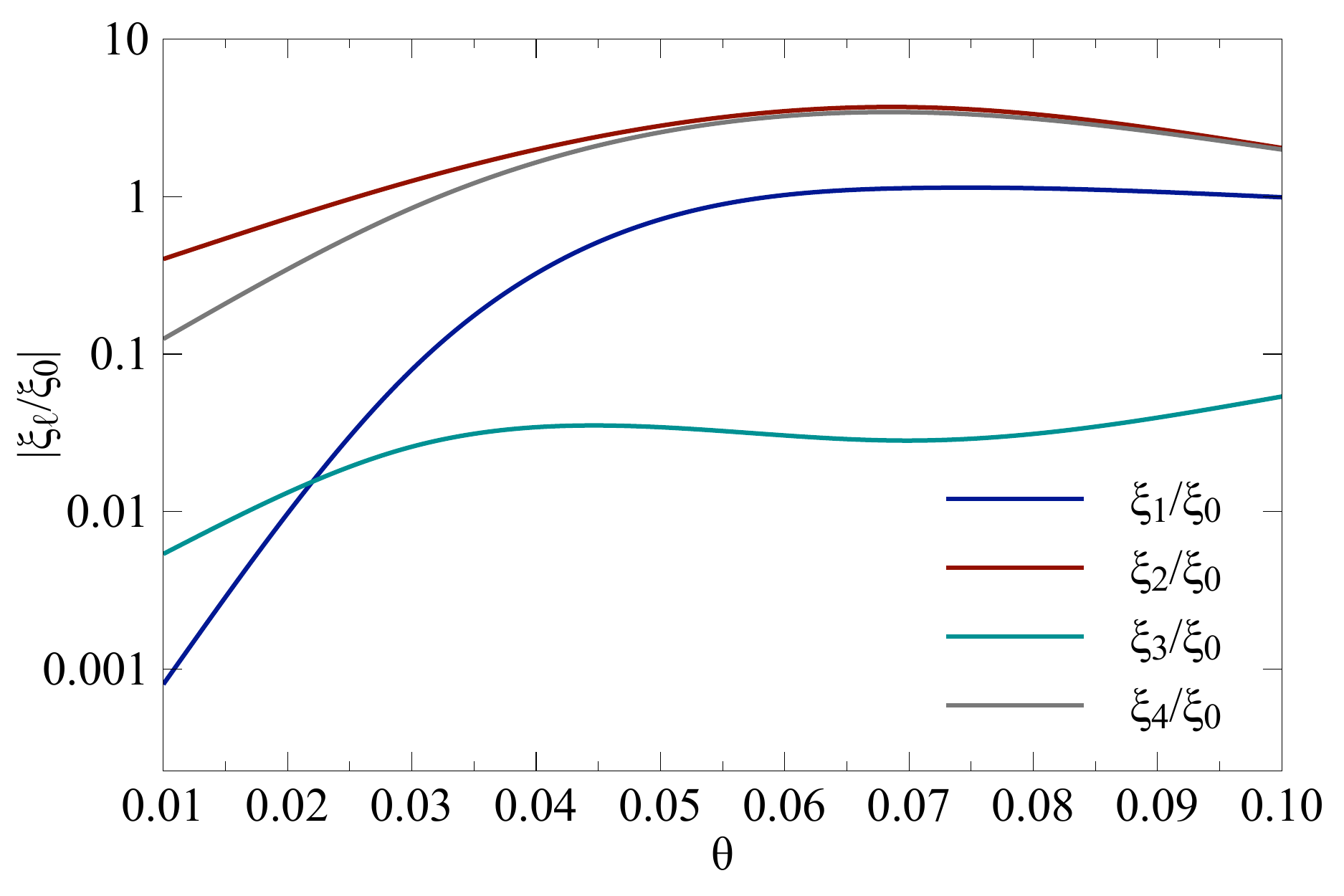}
\includegraphics[width=0.47\linewidth]{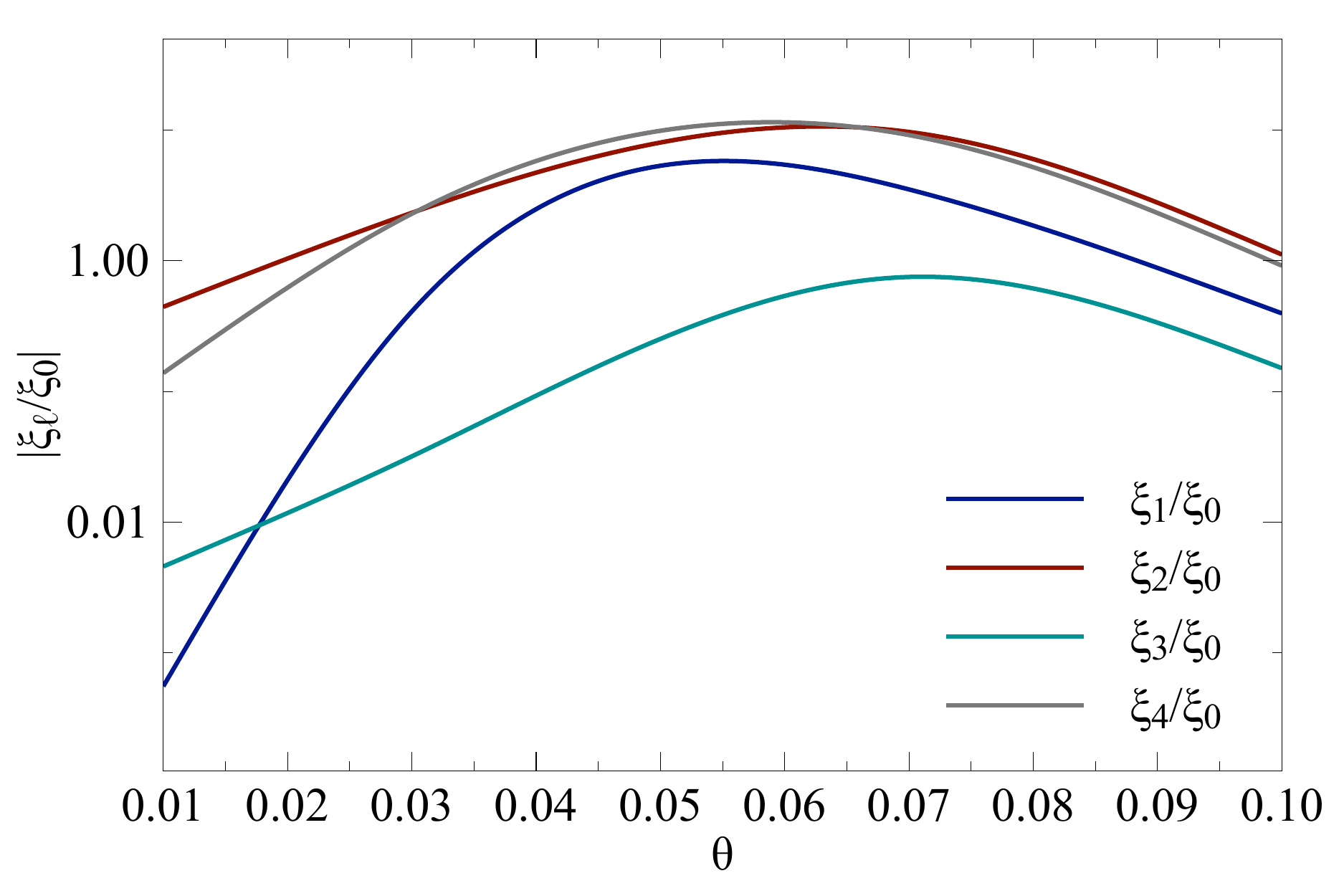}
\caption{Absolute value of multipoles of the correlation function relative to the monopole, as a function of the separation angle $\theta$, for different values of $z_2$. {\it Top left:} $z_2=0.3$, {\it top right:} $z_2=0.5$, {\it bottom left:} $z_2=1$ and {\it bottom right:} $z_2=2$.}
\label{fig:multip_G}
\end{figure}

In Figure~\ref{fig:multip_G} we show the absolute value of the five multipoles as a function of the angular separation $\theta$, for different values of $z_2$.
As expected, the odd multipoles are considerably smaller than the even ones for small angular separation and at low redshift. However, their amplitude can grow to be comparable to the even ones at large angular separations and for moderate to high redshift.

\section{Multipolar non-Gaussianity }
\label{sec:nG_relativistic }

On very large scales, both primordial non-Gaussianity and relativistic effects have a growing impact on clustering. A comparison of the two effects in the angular power spectrum $C_\ell $ has been made by~\cite{Bruni:2011ta, Maartens:2012rh}, including integrated relativistic effects but fixing the redshift. Using the angular power spectrum avoids the flat-sky approximation.  The alternative is to use the power spectrum $P(k)$ at fixed redshift, as in~\cite{Jeong:2011as, Yoo:2012se}, which neglect integrated effects. This approach implicitly uses the flat-sky approximation. 
Here we generalize previous results to allow varying redshift. We include all geometric and local relativistic effects by investigating the multipoles -- both even and odd -- of the correlation function. 

\begin{figure}[!htbp]
\includegraphics[width=0.47\linewidth]{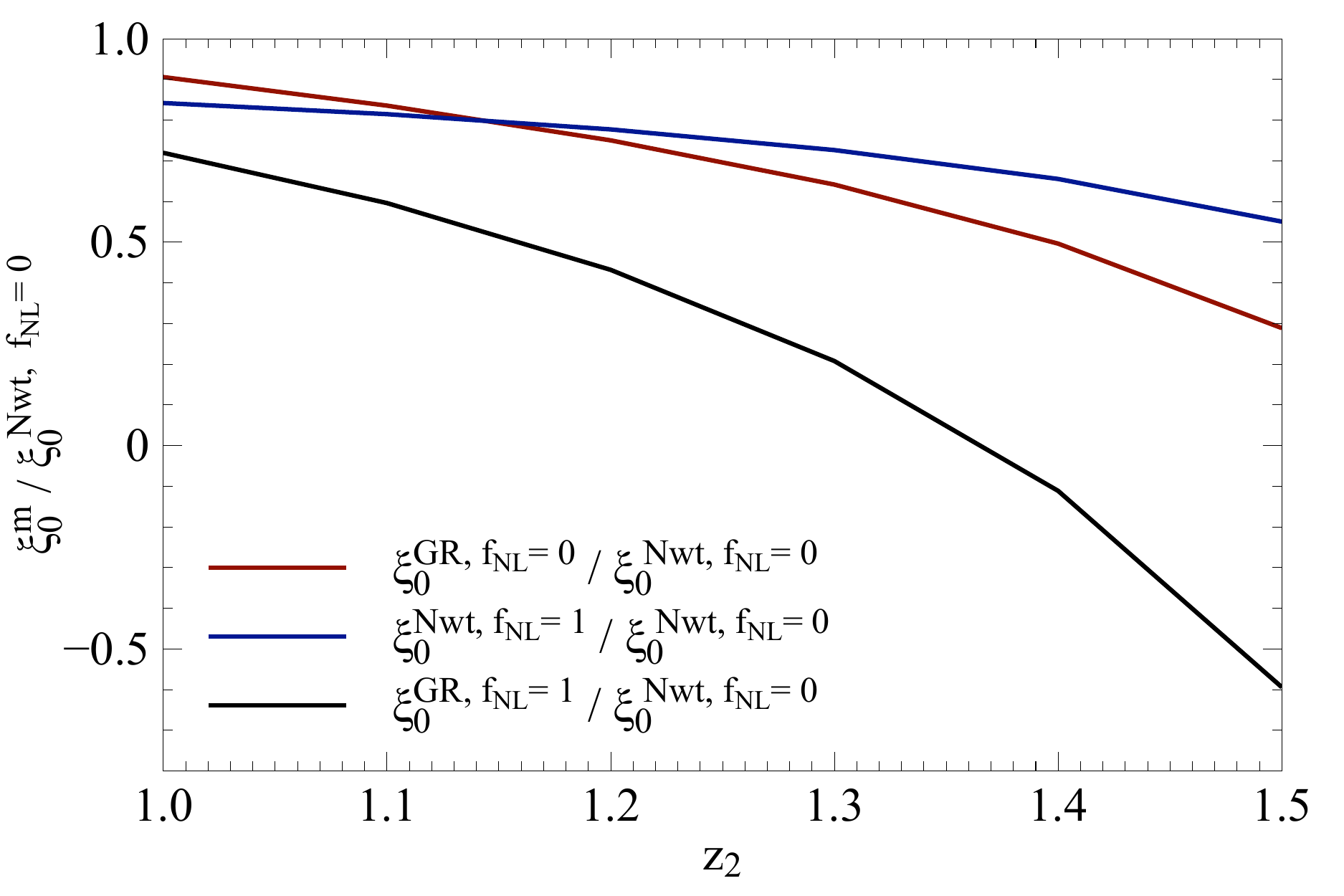}
\caption{Corrections to the Newtonian Gaussian monopole of the correlation function, due to: non-Gaussianity with $f_{\rm NL}=1$ (blue line);   relativistic  corrections (red line);  and the combination of the two effects (black line).  The angular separation is  $\theta=0.1$}
\label{fig:xi0warelativistic fnl}
\end{figure}

\begin{figure}[!htbp]
\includegraphics[width=0.47\linewidth]{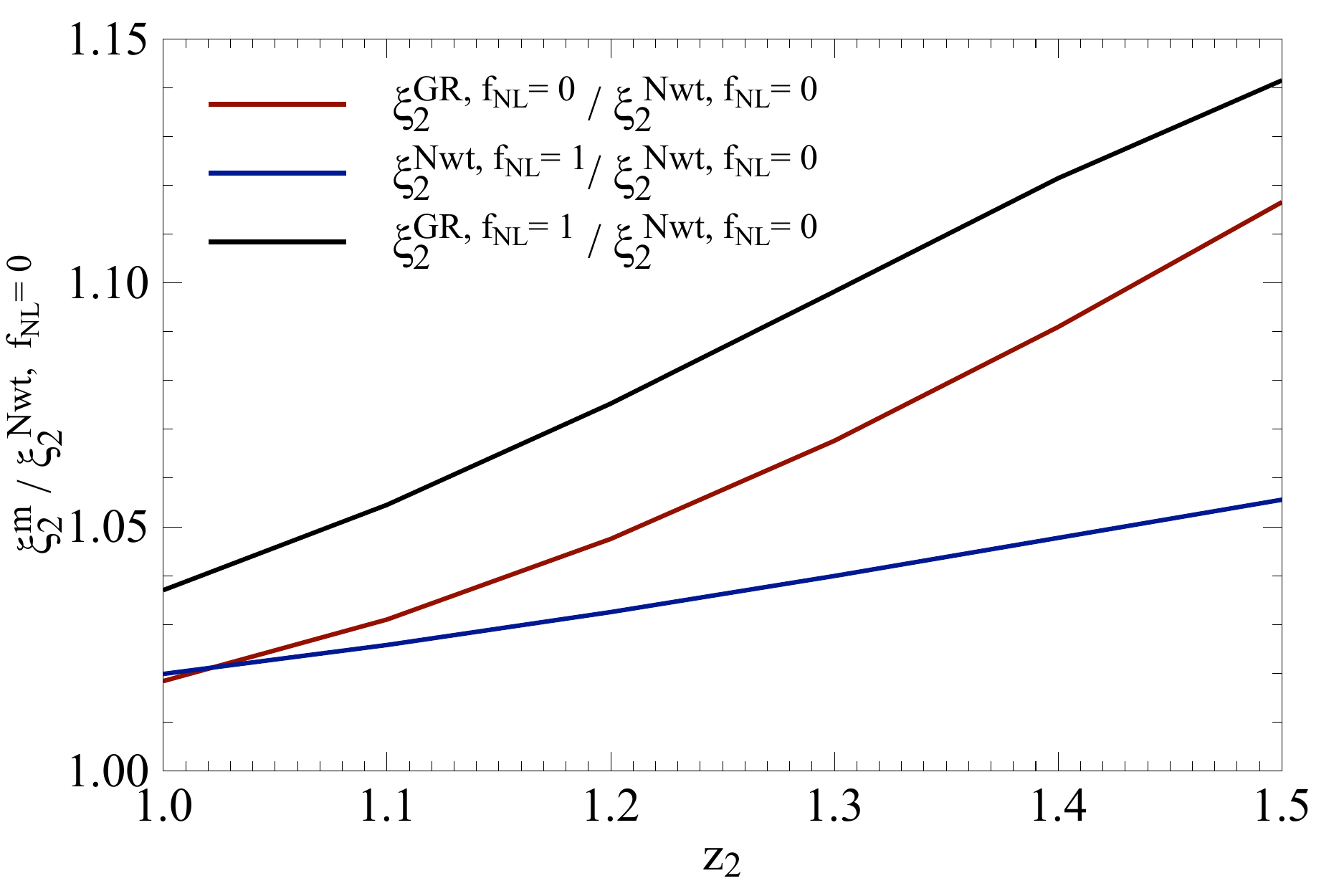}
\includegraphics[width=0.47\linewidth]{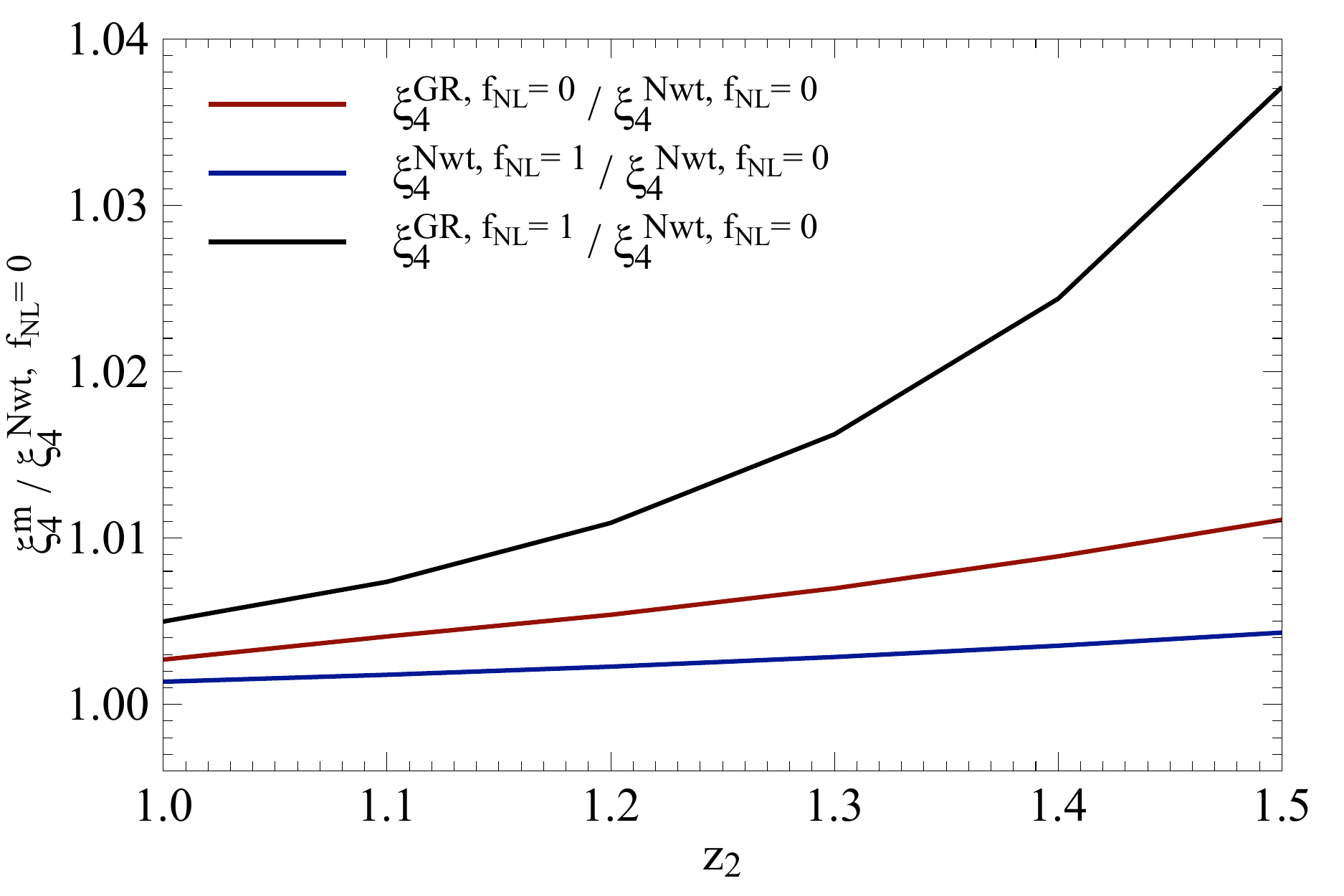}
\caption{As in Figure \ref{fig:xi0warelativistic fnl}, for the quadrupole ({\em left}) and hexadecapole ({\em right}).}
\label{fig:xi24warelativistic fnl}
\end{figure}

\begin{figure}[!htbp]
\includegraphics[width=0.47\linewidth]{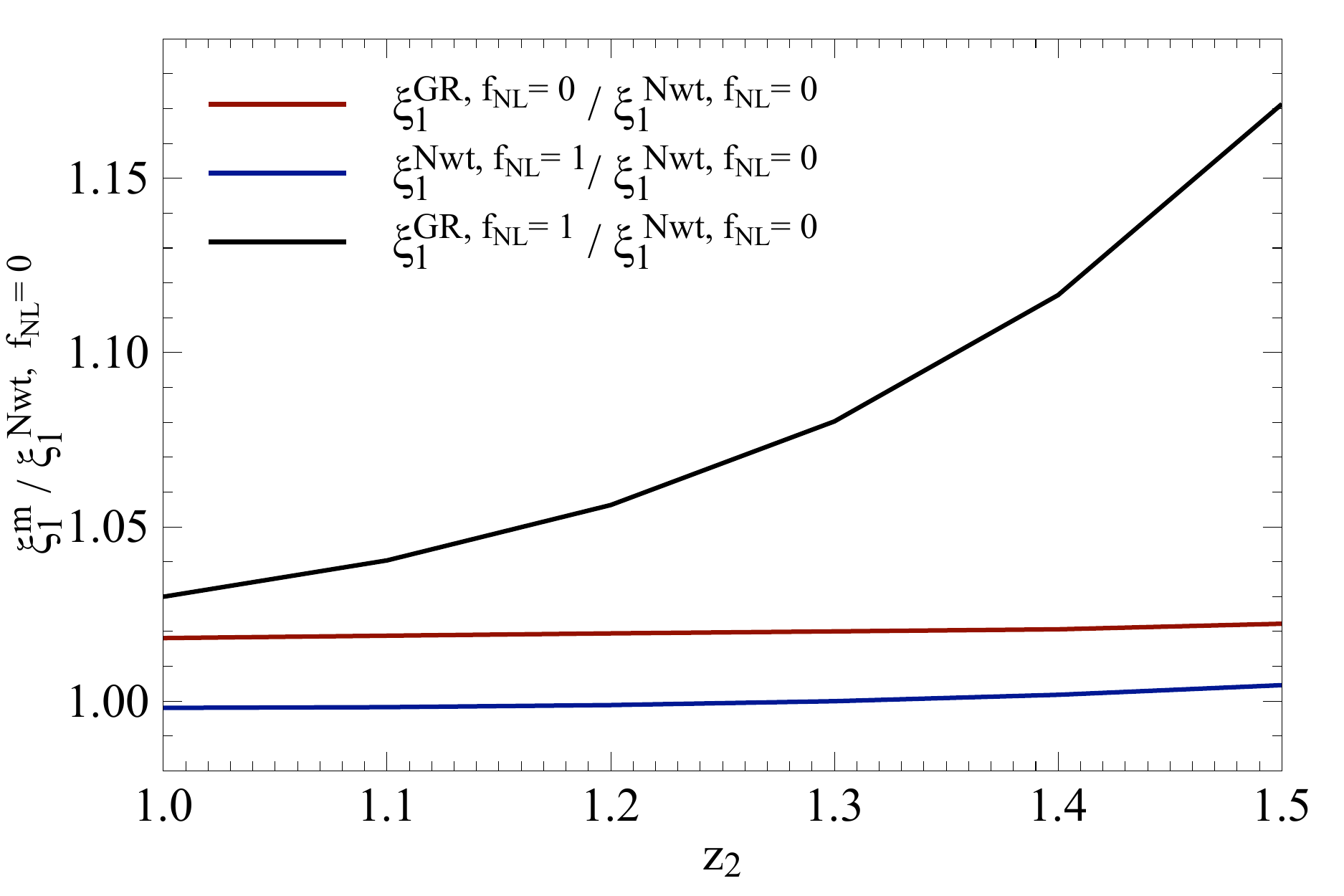}
\includegraphics[width=0.47\linewidth]{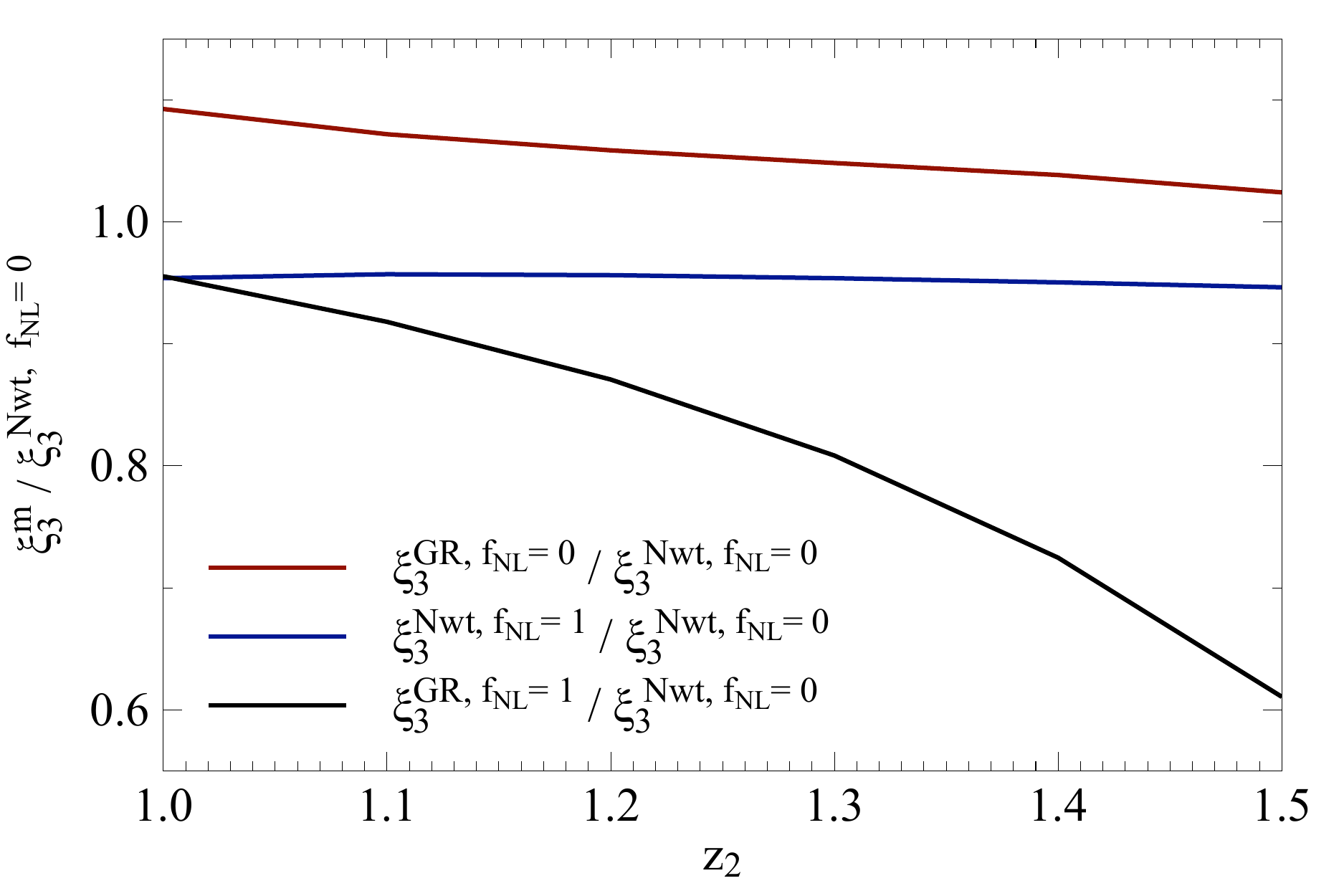}
\caption{As in Figure \ref{fig:xi0warelativistic fnl}, for the dipole ({\em left}) and octupole ({\em right}).}
\label{fig:xi13warelativistic fnl}
\end{figure}

The results are shown
in Figures~\ref{fig:xi0warelativistic fnl}, \ref{fig:xi24warelativistic fnl} and \ref{fig:xi13warelativistic fnl}, using a large angular separation. 
We normalize to the Newtonian Gaussian multipoles $\xi_L^{{\rm Nwt,}\,f_{\rm NL}=0}$, and consider the 3 cases:
\begin{equation}
\mbox{GR effects,}~ f_{\rm NL}=0;~~ \mbox{Newtonian,}~ f_{\rm NL}=1;~~ \mbox{GR effects,}~ f_{\rm NL}=1.
\end{equation}
In line with the recent Planck results we use a small value of $f_{\rm NL} =1$.

The very large scales we use here are not probed by current surveys, but they might be in the future, such as in PFS- and Euclid-like surveys. The exact details and significance of those measurements will depend on specific surveys and the precision it will be possible to obtain.
Our results indicate that
relativistic and non-Gaussian (consistent with Planck) effects are comparable, and that looking at the 3D structure by using all the different multipoles can help to distinguish between the two effects. Once again, the additional information coming from the odd multipoles is particularly useful.

\section{Modified gravity and non-Gaussianity}
\label{sec:growth}

As can be seen from Equations~\eqref{eq:beta}, \eqref{eq:gammaz} and \eqref{eq:ng-bias}, the effects of primordial non-Gaussianity and relativistic corrections both depend on the growth factor $f$.
Modified gravity is an alternative to dark energy for driving the late-time acceleration of the Universe. Most modified gravity theories are relativistic, and so the observable galaxy overdensity acquires the same form of relativistic corrections as in the case of GR. Primordial non-Gaussianity will also produce the same enhanced bias on large scales. 

A detailed analysis of the effect of modified gravity on large-scale corrections to the correlation function is beyond the scope of this paper and requires a careful treatment, so we leave it to a future work. Instead, we use a simple modification of the growth rate to illustrate what these effects could include. We parametrize the  growth factor using the formula suggested in~\cite{Linder:2005in}:
\begin{equation}
\label{eq:gamma}
f(z) = \Omega_m(z)^{\gamma} \; .
\end{equation}
The parameter $\gamma$ (not to be confused with the relativistic function defined in Equation~\eqref{eq:gammaz}) is approximately constant in a variety of models, and can be used to constrain models of gravity~\cite{Raccanelli:2012gt}. Its value is different for different theories. For $\Lambda$CDM and for simple dynamical dark energy models in GR,   $\gamma\approx 0.55$, while $\gamma\approx0.68$ in the brane-world DGP theory. For other theories,  such as Galileon theories~\cite{Kobayashi:2010wa}, it is scale- and redshift-dependent,.

\begin{figure}[!htbp]
\includegraphics[width=0.47\linewidth]{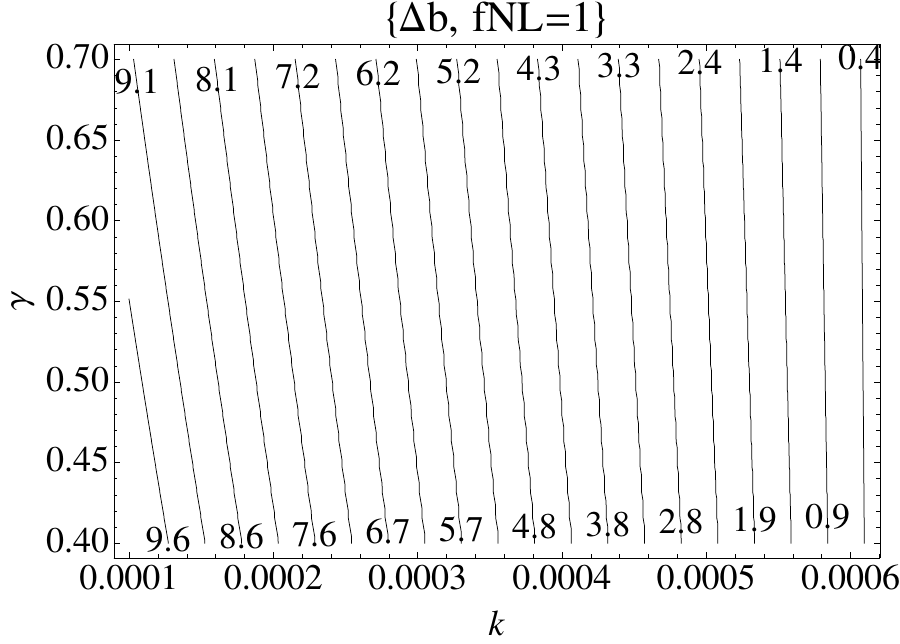}
\includegraphics[width=0.47\linewidth]{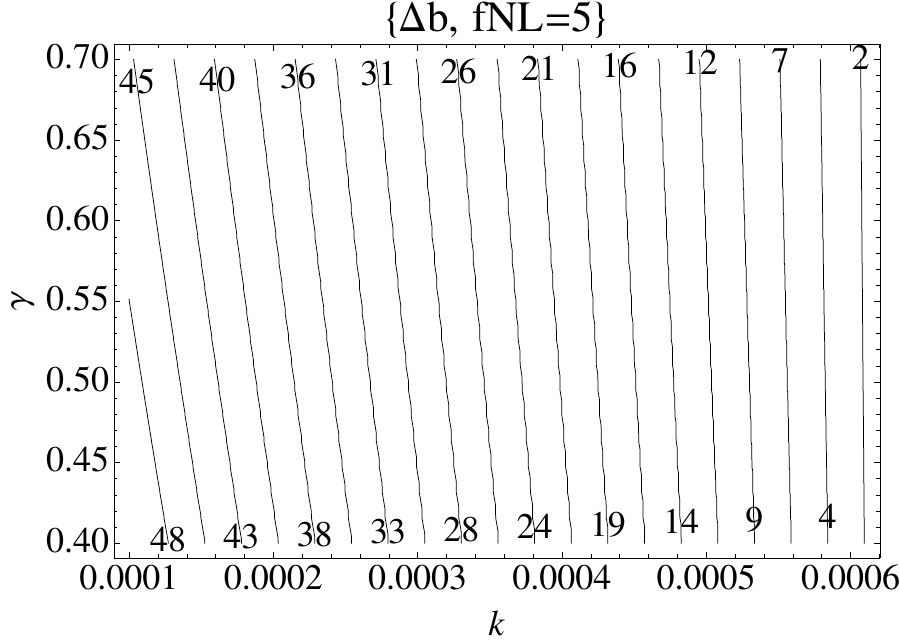}
\includegraphics[width=0.47\linewidth]{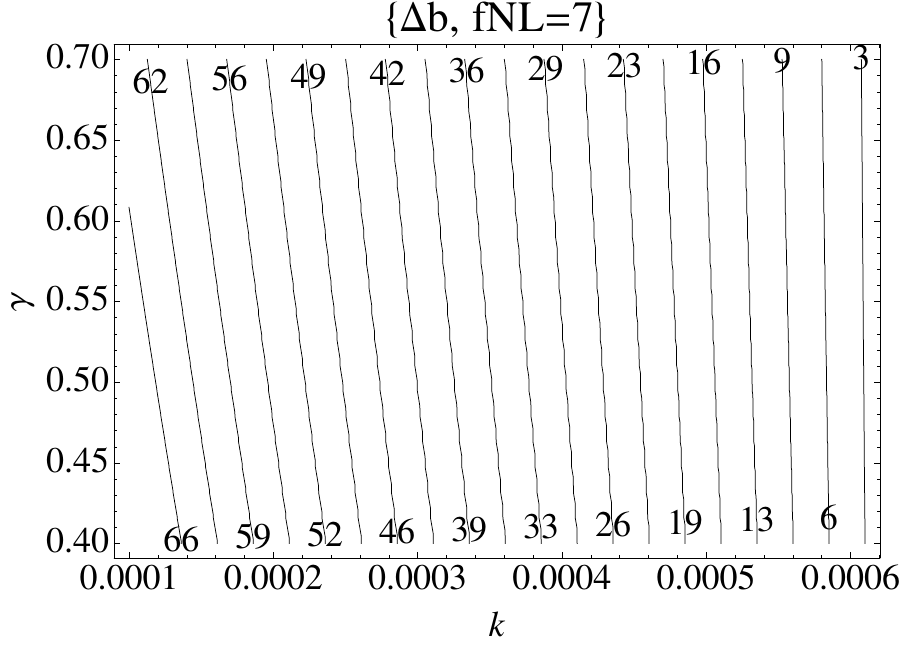}
\includegraphics[width=0.47\linewidth]{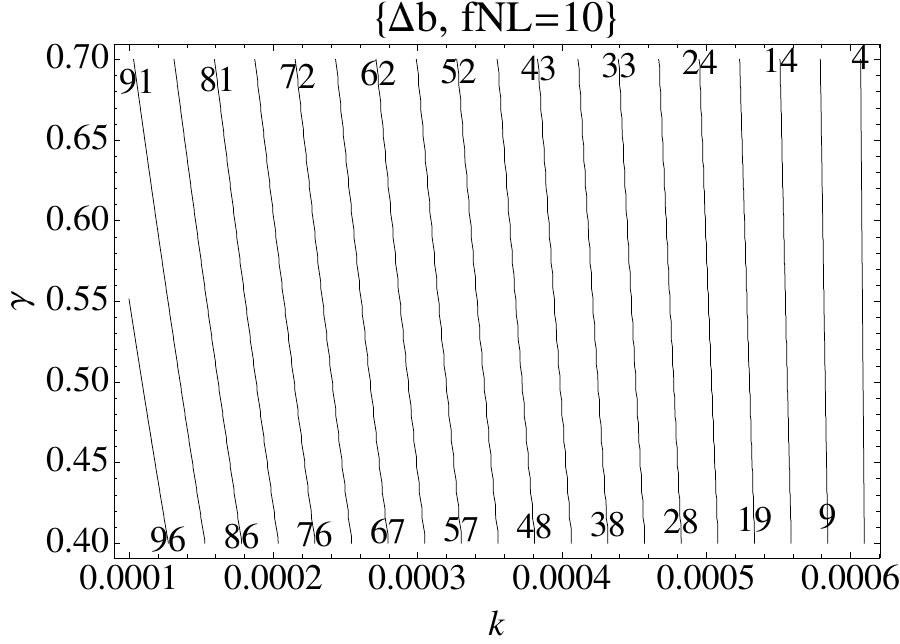}
\caption{Effect of modified growth on the non-Gaussian correction to the bias $\Delta b$, for different values of \fnlt. Lines show the value of $\Delta b$ for the corresponding scale $k$ and growth rate $\gamma$.}
\label{fig:growth_fnl}
\end{figure}

In Figure~\ref{fig:growth_fnl} we show what happens to the non-Gaussian correction to the bias when one modifies the growth rate. A larger growth parameter $\gamma$ will decrease $\Delta b$, and the plots show this effect:  an enhanced (reduced) growth  decreases (increases) the effect of a positive primordial non-Gaussianity. 

\begin{figure}[!htbp]
\includegraphics[width=0.47\linewidth]{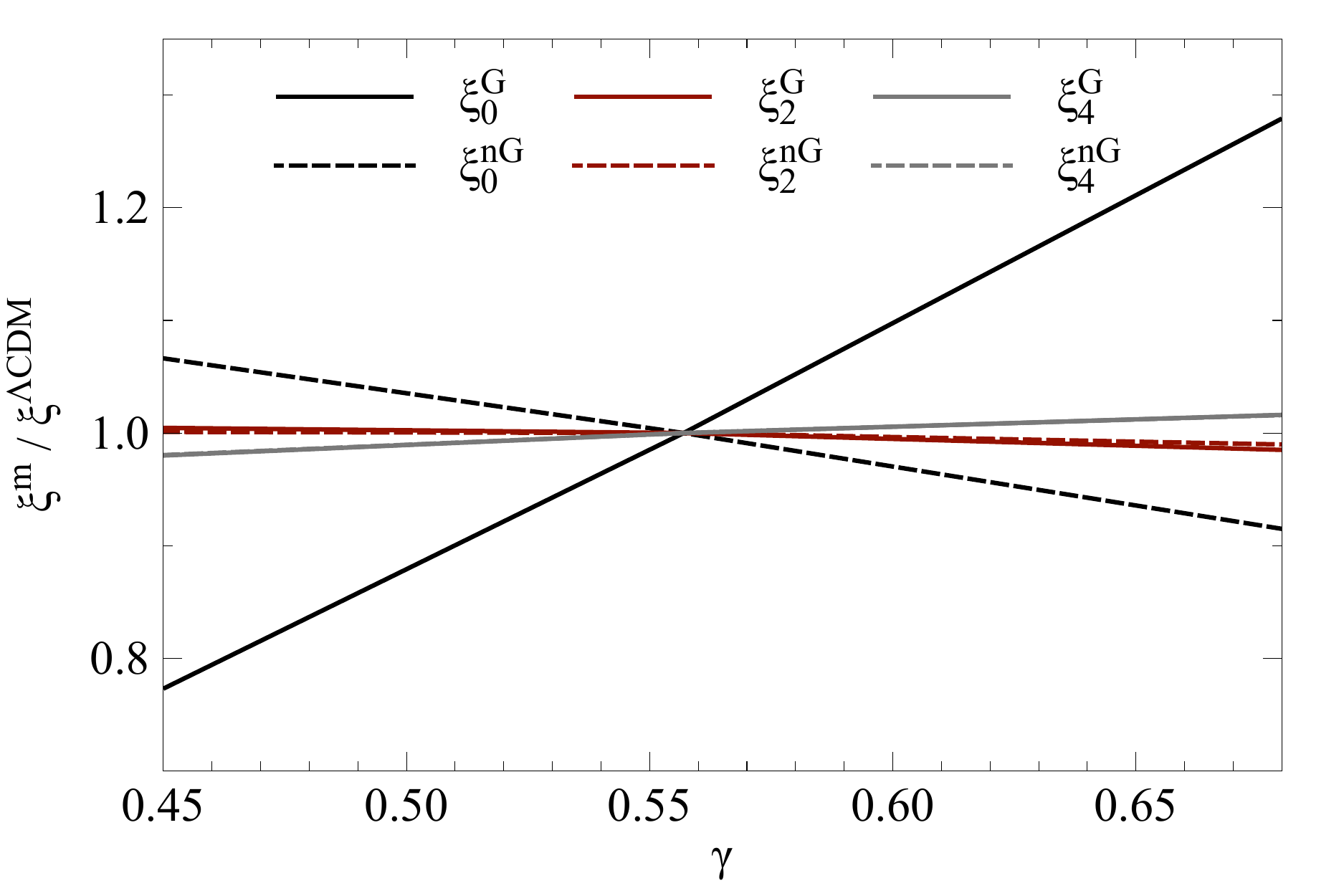}
\includegraphics[width=0.47\linewidth]{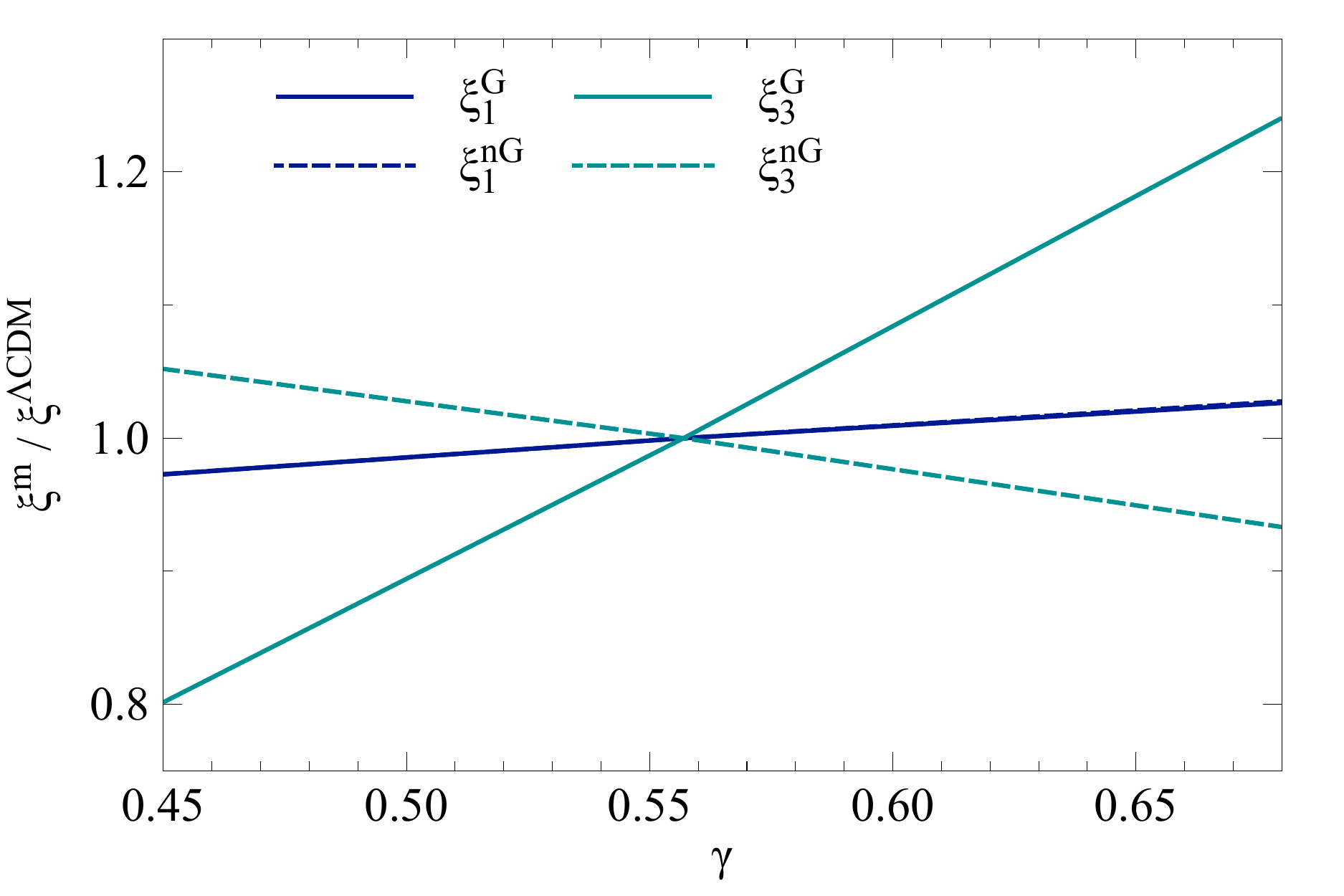}
\caption{Effect of modified growth on the multipoles of the correlation function.
{\it Left Panel:} even multipoles, {\it Right Panel:} odd multipoles. The correlation function with modified growth $\xi^{\rm m}$, relative to the relativistic $\Lambda$CDM correlation function, is shown in the Gaussian (solid lines) and \fnlt=10 (dashed lines) cases.  Here $ z=1$ and $\theta=0.1$.}
\label{fig:xigrowth}
\end{figure}

In Figure~\ref{fig:xigrowth} we show an example of the impact of modified growth on the correlation function with relativistic  effects included, in the Gaussian and non-Gaussian cases. Here we compute the multipoles as in Equation~\eqref{eq:multip_standard}, and we use a stronger non-Gaussianity $f_{\rm NL}=10$ to show up the difference from the Guassian case more clearly. We see that modified growth and non-Gaussianity have different effects on different multipoles, with the 
monopole
and octupole being in this case very sensitive, while the even multipoles are less sensitive to these modifications. This once again emphasizes the importance of taking into account the fact that odd multipoles are nonzero in the 3D case.

It is worth noting that effects of non-Gaussianity and modified growth on different multipoles are in general redshift- and angle-dependent, so this is just an illustrative example, and a more detailed analysis is left to future work.

\section{Conclusions}
\label{sec:conclusions}

The clustering of galaxies is one of the most important tools for present and forthcoming cosmological surveys.
Until now most analyses have been performed using the flat-sky and/or Newtonian approximations. This is adequate for past and present surveys, which analyze galaxy clustering on scales well below the horizon, but future wide and deep surveys will need to utilize a more precise modeling, including all geometry and relativistic corrections.
Future surveys will probe large cosmological volumes and measure the properties of the galaxy field on scales comparable to the horizon, where effects of modified gravity and signatures from the very early Universe will be more prominent.
This is going to be important in particular for measuring primordial non-Gaussianity, which grows as $k^{-2}$ on very large scales. 
In addition, relativistic corrections to the Newtonian galaxy overdensity also grow on large scales. 
For this reason, a non-relativistic approach would be insufficient for precise large-scales measurements of non-Gaussianity.

In this paper we studied the modeling of the 2-point correlation function of galaxies at very large scales using an accurate treatment of the geometry of the system and of effects due to relativistic corrections. We built on previous work in the Newtonian approximation~\cite{Szalay:1997cc, Matsubara:1999du, Szapudi:2004gh, Papai:2008bd, Raccanelli:2010hk, Samushia:2011cs}, and its relativistic generalization~\cite{Bertacca:2012tp}.
The full 3D description of the correlation function, including wide-angle, radial and relativistic corrections  introduces some new features that have not been studied before.

In particular, we analyzed the Legendre polynomial expansion of the local part of the correlation function, and showed that, while the odd Legendre multipoles vanish in the Kaiser approximation, when considering carefully the geometry of the system, they are in general non-zero, potentially allowing us to extract more information about the structure of galaxy clustering.
We also showed that, when looking at the 3D structure of the correlation function, one can more easily distinguish between corrections induced by relativistic effects and by primordial non-Gaussianity. 

Finally, we investigated how non-Gaussianity depends on the modified gravity theories, using a simple parametrization for the growth rate. We showed how the Legendre multipoles are affected by such modifications in a heuristic manner.

It appears that having additional angular information about the clustering of structures on large scales will lead to more precise analyses of the behavior of gravity on those scales. 
In a companion paper~\cite{Raccanelliradial} we will present a detailed analysis of the local and the integrated relativistic effects,
and in a following paper we will investigate the measurability of the odd multipoles in current and future surveys. This will most likely be important for future spectroscopic surveys that aim to probe the clustering of galaxies at very large scales with high precision. Our study indicates that they will be rich in new physical signals that require accurate modeling.

\acknowledgements{We thank Nicola Bartolo, Kazuya Koyama, Marc Manera, Sabino Matarrese, Will Percival, Ashley Ross, Lado Samushia and Gianmassimo Tasinato for helpful discussions. Part of the research described in this paper was carried out at the Jet Propulsion Laboratory, California Institute of Technology, under a contract with the National Aeronautics and Space Administration. The work of DB and RM was supported by the South African Square Kilometre Array Project and the South African National Research Foundation. RM was also supported by the UK Science \& Technology Facilities Council (grant nos. ST/H002774/1 and ST/K0090X/1). 
}

\appendix

\section{Alternative multipole expansion}

Here we derive an alternative expression for the correlation function \txiss as the combination of two Legendre polynomials (instead of the $S_{\ell_1\ell_2L}$ of Equation~\eqref{eq:xi_ss}) -- one for the angular dependence of \tphi and one for $\theta$. This method appears to be more natural for looking at the behaviour of the  multipoles. 
Using the fact that the terms proportional to $\sin^{2n} =(1-\cos^{2})^n$ can be rewritten as a Legendre polynomial, we obtain a new decomposition (for simplicity we set $f_{\rm NL} = 0$):
\begin{equation}
\label{xisum}
\xi (z_2, \theta, \varphi) = b(z_1) b(z_2) \sum_{\tilde{L} , \tilde{\ell}} \; \Psi_{\tilde{L} \tilde{\ell}} (z_2, \theta, \varphi) \; \mathcal{P}_{\tilde{L} }(\cos\varphi) \; \mathcal{P}_{ \tilde{\ell}}(\cos\theta) \; ,
\end{equation}
where now $z_1=z_1(z_2, \theta, \varphi)$. The coefficients $\Psi_{\tilde{L} \tilde{\ell}}$ are determined by $\xi_L^n$, and are given in Appendix B.
If we set $\alpha = \gamma = \theta = 0$, then we remove the wide-angle and mode coupling effects from $\alpha,\theta$ and the relativistic corrections from $\gamma$. If we also set $z_1 \simeq z_2$, and resum over the Legendre polynomials of $\theta$, then we recover the Kaiser expressions for the multipoles (see e.g.~\cite{Hamilton:1997zq}):
\begin{eqnarray}
{\xi \over b^2} &=&  \left[1+\frac{2}{3}\beta + \frac{1}{5}\beta^2 \right] \,\xi_0^{0} \,
    \mathcal{P}_{0}(\cos\varphi) - 
    \left[\frac{4}{3}\beta^2 + \frac{4}{7}\beta^2\right] \, \xi_2^{0} \, \mathcal{P}_{2}(\cos\varphi)
   + \frac{8}{35} \beta^2 \, \xi_4^{0} \, \mathcal{P}_{4}(\cos\varphi).\ \label{ppxiss}
 \end{eqnarray}
%


In Eq. (\ref{xisum}),
note that the coefficients $\Psi_{\tilde L \tilde\ell}$, unlike in the plane-parallel case, depend on the arguments of the polynomials. In this case the usual Legendre polynomial is not a correct basis for the expansion of  the correlation function $\xi$.

A possible alternative approach to obtain the 3D relativistic equivalent of the usual multipoles could be to  re-sum terms with the same $L$ in Equation~\eqref{eq:multip_standard}, using Equation~\eqref{xisum}:
\begin{eqnarray}
\label{eq:multip}
\tilde{\xi}_{L} = \frac{2 L +1}{2} \frac{\pi}{\pi-2\theta} \int_{\theta}^{\pi-\theta} ~ {\rm d} \, \varphi ~   \mathcal{P}_L \left\{\cos\left[\frac{\pi \left( \varphi - \theta \right)}{\pi - 2\theta}\right]\right\}\, \sin\left[\frac{\pi \left( \varphi - \theta \right)}{\pi - 2\theta}\right] b_1b_2  \sum_\ell \left[ \Psi_{L \ell} \mathcal{P}_L(\cos\varphi) \mathcal{P}_\ell(\cos\theta)\right].
\end{eqnarray}

However, there are two important points that explain why we cannot use Equation \eqref{eq:multip}:
\begin{itemize}
\item[(1)] The new integration range over $\varphi$  breaks the orthogonality, as shown by Eq. (\ref{ort-legendreNEW}).

%

\item[(2)] The coefficients  $\Psi_{L \ell}$, $\xi_L^n$ and $b_1$ depend on  $\varphi$  through $z_1$ and $\chi_{12}$. As a consequence, in the 3D formalism,
the $\varphi$ dependence of the power spectrum in Equation~\eqref{eq:xiLn} destroys the orthogonality of Legendre polynomials. 
\end{itemize}

\begin{figure}
\includegraphics[width=0.7\linewidth]{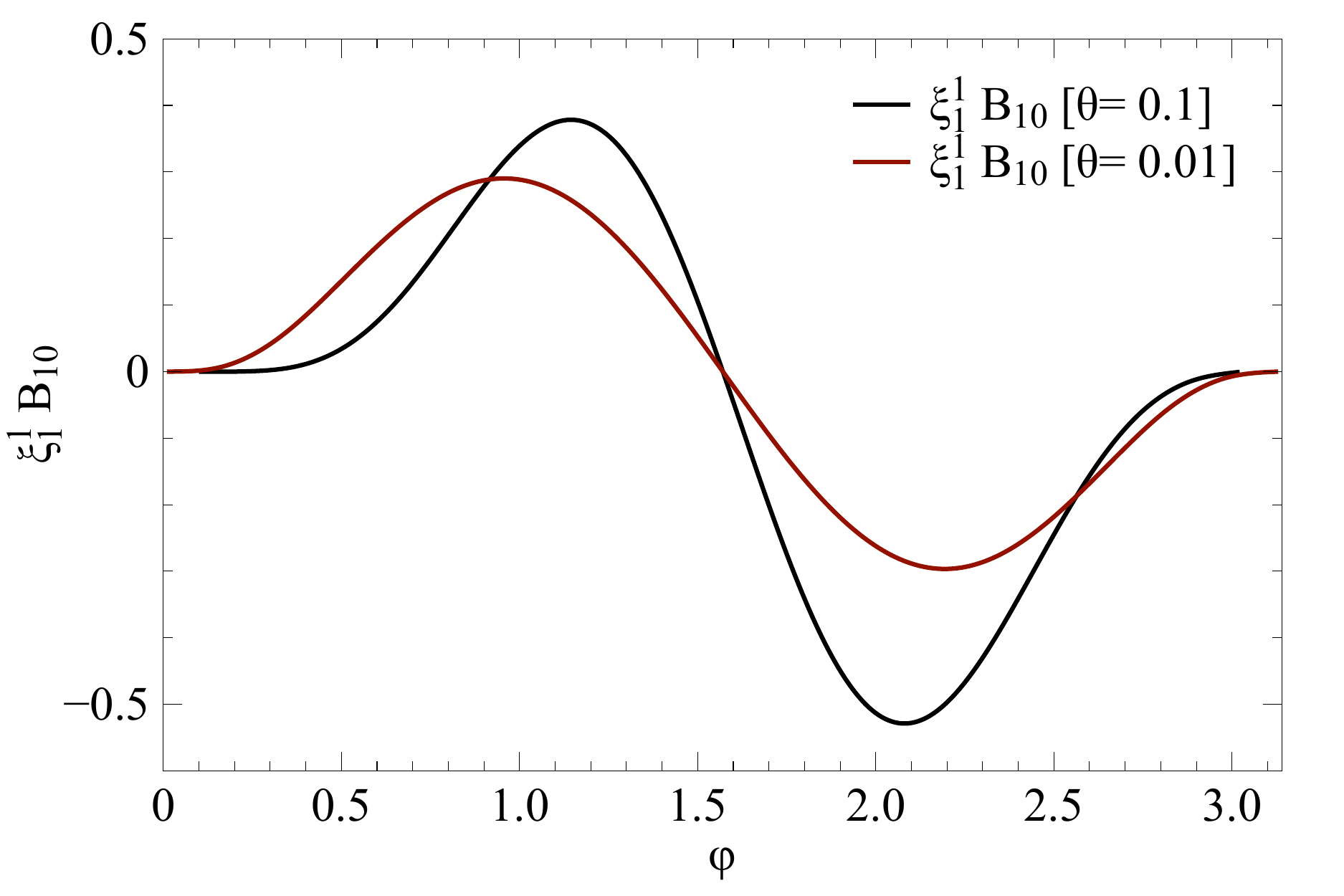}
\caption{$\xi_1^1 B_{10}$ as a function of $\varphi$, for two values of \ttheta.
}
\label{fig:xi11ort}
\end{figure}

We can illustrate these points via an example.
In Figure~\ref{fig:xi11ort} we plot $ \xi_\ell^n B_{\ell L}$, for two different values of $\theta$, where:
\begin{eqnarray}
\label{eq:xiB}
B_{\ell L}(\theta, \varphi) &\equiv& \frac{2L+1}{2} \frac{\pi}{\pi-2\theta} ~ \mathcal{P}_\ell \left(\cos\varphi\right) ~\mathcal{P}_\ell \left(\cos\theta\right) ~\mathcal{P}_L \left\{\cos \left[ \frac{\pi(\varphi-\theta)}{\pi - 2\theta}\right]\right\} ~\sin \left[\frac{\pi(\varphi-\theta)}{\pi - 2\theta}\right] \,.
\end{eqnarray}

The terms in the sum over $\ell$ in the integrand in Equation \eqref{eq:multip} are proportional to $\xi_\ell^n B_{\ell L}$. Figure~\ref{fig:xi11ort}  shows that for $\theta$ not small, these terms lose the symmetry about $\varphi=\pi/2$. Therefore the integral over $\varphi$ will be nonzero, destroying orthogonality.
For small $\theta$, Figure~\ref{fig:xi11ort}  shows that we begin to recover the symmetry of the flat-sky case, leading to a zero integral over $\varphi$.
It is also interesting to note that the integrand goes always to zero for $\varphi = \pi/2$, because this is the case when the two galaxies are at the same $z$, and so the problem is again symmetric. 
For pairs almost transverse to the line of sight, one could indeed use the formalism of Equation~\eqref{eq:multip} to more easily compute the multipoles.


In conclusion, 
if the usual orthogonal properties were respected -- i.e. if $\Psi_{\tilde L \ell}$ did not depend on $\theta$ and $\varphi$ and within the integral we had $\mathcal P_L (\cos \varphi) \sin \phi$ -- then Eq. (\ref{eq:multip}) would be correct. In this case, $\tilde{L}= L$ and we sum only over $\ell$. Unfortunately, 
due to the 3D properties of $\xi$ this is not true (unless $\theta$ goes to zero) and thus we can only obtain the projection in multipoles of $\xi$ numerically. For these reasons, ${\xi}_{L} \neq \tilde{\xi}_{L}$ and we need to evaluate the multipole expansion over $\varphi$ using Equation~\eqref{eq:multip_standard}.

\section{ The coefficients $\Psi_{\tilde{L} \tilde{\ell}}$ }

\begin{eqnarray}
\label{eq:psis}
\Psi_{00}&=&\left(1+\frac{\beta_1}{3}+\frac{\beta_2}{3} + \frac{29}{225} \beta_1 \beta_2 \right) \xi_0^0 -\left(\gamma_1+\gamma_2+\frac{1}{3}\beta_1\gamma_2+\frac{1}{3}\gamma_1\beta_2+\frac{1}{9}\beta_1\beta_2\frac{\alpha_1}{\chi_1}\frac{\alpha_2}{\chi_2}\right) \xi^2_0 + \nonumber \\
&&+ \gamma_1 \gamma_2 \xi^4_0+\sin(\varphi)\sin(\theta) \left[\left(1+\frac{1}{3}\beta_1\right)\beta_2\frac{\alpha_2}{\chi_2}+\left(1+\frac{1}{3}\beta_2\right)\beta_1\frac{\alpha_1}{\chi_1}\right] \xi_1^1 + \nonumber \\ 
&&-\sin(\varphi)\sin(\theta)\left(\gamma_1 \beta_2\frac{\alpha_2}{\chi_2} +\beta_1\frac{\alpha_1}{\chi_1} \gamma_2 \right)\xi^3_1 
-\left(\frac{2}{9}\beta_1+\frac{2}{9}\beta_2+\frac{44}{315}\beta_1\beta_2\right)\xi^0_2 + 
\nonumber \\ 
&&+\frac{2}{9}\left(\beta_1\beta_2\frac{\alpha_1}{\chi_1}\frac{\alpha_2}{\chi_2}+\gamma_1\beta_2+\beta_1 \gamma_2\right)\xi^2_2 + \frac{32}{1575}\beta_1 \beta_2 \xi^0_4 \;, \nonumber \\ \nonumber\\
 \Psi_{11}&=& 
\left[-\left(1+\frac{7}{25}\beta_1\right)\beta_2\frac{\alpha_2}{\chi_2}+\left(1+\frac{7}{25}\beta_2\right)\beta_1\frac{\alpha_1}{\chi_1}\right] \xi_1^1 + \left( \gamma_1 \beta_2\frac{\alpha_2}{\chi_2} - \beta_1\frac{\alpha_1}{\chi_1}  \gamma_2 \right) \xi^3_1  + \nonumber\\
&&+2 \sin(\varphi)\sin(\theta)\left(\beta_2-\beta_1\right)\xi^0_2 + 2 \sin(\varphi)\sin(\theta)\left(\beta_1\gamma_2-\gamma_1\beta_2\right)\xi^2_2 + \nonumber\\
&& + \frac{2}{25}\left( \beta_1\frac{\alpha_1}{\chi_1}\beta_2-\beta_1 \beta_2\frac{\alpha_2}{\chi_2} \right) \xi^1_3\;, \nonumber \\
   \Psi_{02}&=& - \frac{16}{315} \beta_1 \beta_2 \xi_0^0+ \frac{4}{9}\beta_1\beta_2\frac{\alpha_1}{\chi_1}\frac{\alpha_2}{\chi_2} \xi^2_0 - \frac{8}{15} \sin(\varphi)\sin(\theta)\beta_1\beta_2 \left(\frac{\alpha_1}{\chi_1}+\frac{\alpha_2}{\chi_2}\right) \xi_1^1 + \nonumber \\ 
 &&+ \left(\frac{2}{9}\beta_1+\frac{2}{9}\beta_2+\frac{100}{441}\beta_1\beta_2\right)\xi^0_2  -\frac{2}{9}\left(\beta_1\beta_2\frac{\alpha_1}{\chi_1}\frac{\alpha_2}{\chi_2}+\gamma_1\beta_2+\beta_1 \gamma_2\right)\xi^2_2\; +\nonumber\\
&&  + \frac{2}{15} \sin(\varphi)\sin(\theta)\beta_1\beta_2 \left(\frac{\alpha_1}{\chi_1}+\frac{\alpha_2}{\chi_2}\right) \xi_3^1- \frac{88}{2205}\beta_1 \beta_2 \xi^0_4 \;, \nonumber \\ 
 \Psi_{20} &=& \left(\frac{2}{9}\beta_1+\frac{2}{9}\beta_2+\frac{4}{21}\beta_1\beta_2\right)\xi^0_2 -\frac{2}{9}\left(3\beta_1\beta_2\frac{\alpha_1}{\chi_1}\frac{\alpha_2}{\chi_2}+\gamma_1\beta_2+\beta_1 \gamma_2\right)\xi^2_2\; +\nonumber\\
&& + \frac{2}{3} \sin(\varphi)\sin(\theta)\beta_1\beta_2 \left(\frac{\alpha_1}{\chi_1}+\frac{\alpha_2}{\chi_2}\right) \xi_3^1 - \frac{8}{63}\beta_1 \beta_2 \xi^0_4\;, \nonumber \\
  \Psi_{22} &=& - \left(\frac{8}{9}\beta_1+\frac{8}{9}\beta_2+\frac{16}{21}\beta_1\beta_2\right)\xi^0_2 +\frac{8}{9}\left(\gamma_1\beta_2+\beta_1 \gamma_2\right)\xi^2_2\; +  \nonumber\\
&& + \frac{8}{63}\beta_1 \beta_2 \xi^0_4 \;, \nonumber \\  
 \Psi_{13} &=& \frac{8}{25} \beta_1\beta_2 \left(\frac{\alpha_1}{\chi_1} -\frac{\alpha_2}{\chi_2}\right)  \xi^1_1  -  \frac{2}{25} \beta_1\beta_2 \left( \frac{\alpha_1}{\chi_1} -\frac{\alpha_2}{\chi_2}\right)  \xi^1_3\;, \nonumber \\
 \Psi_{31} &=& - \frac{2}{5} \beta_1\beta_2 \left(\frac{\alpha_1}{\chi_1} -\frac{\alpha_2}{\chi_2}\right)  \xi^1_3\;, \nonumber \\
 \Psi_{04} &=& \frac{64}{525}\beta_1 \beta_2 \xi^0_0 - \frac{64}{735}\beta_1 \beta_2 \xi^0_2 + \frac{24}{1225}\beta_1 \beta_2 \xi^0_4\;, \nonumber \\
 \Psi_{40} &=& \frac{8}{35}\beta_1 \beta_2 \xi^0_4 \; , \nonumber
\end{eqnarray}
where a subscript $i$ denotes evaluation at $z_i$. These coefficients were obtained assuming $f_{\rm NL}=0$; in the non-Gaussian case it is necessary to include the modification to the bias as in Equation~\eqref{eq:ng-bias}.


\begin{thebibliography}{99}

\bibitem{pfs}
	R.~Ellis, {\it et al.} [PFS Team],
	(2012) arXiv:1206.0737.

\bibitem{Amendola:2012ys}
  L.~Amendola {\it et al.}  [Euclid Theory Working Group Collaboration],
  Living Rev.\ Rel.\  {\bf 16} (2013) 6
  [arXiv:1206.1225].

\bibitem{Samushia:2011cs}
  L.~Samushia, W.~J.~Percival and A.~Raccanelli,
  Mon.\ Not.\ Roy.\ Astron.\ Soc.\  {\bf 420} (2012) 2102
  [arXiv:1102.1014].
  
  
  \bibitem{samushiaboss}
L.~Samushia, B.~A.~Reid, M.~White, W.~J.~Percival, A.~J.~Cuesta, L.~Lombriser, M.~Manera and R.~C.~Nichol {\it et al.}, 
Mon.\ Not.\ Roy.\ Astron.\ Soc.\  {\bf 429} (2013) 1514 
[arXiv:1206.5309].

  
  \bibitem{riedboss}
B.~A.~Reid, L.~Samushia, M.~White, W.~J.~Percival, M.~Manera, N.~Padmanabhan, A.~J.~Ross and A.~G.~Sanchez {\it et al.},
Mon.\ Not.\ Roy.\ Astron.\ Soc.\  {\bf 426} (2012) 2719R
[arXiv:1203.6641].
  

\bibitem{sanchezboss}
A.~G.~ Sanchez {\it et al.},
Mon.\ Not.\ Roy.\ Astron.\ Soc.\  {\bf 425} (2012) 415S
[arXiv:1203.6616].

\bibitem{delatorrevipers}
S.~de la Torre, L.~Guzzo, J.~A.~Peacock, E.~Branchini, A.~Iovino, B.~R.~Granett, U.~Abbas and C.~Adami {\it et al.} (2013)
arXiv:1303.2622.


\bibitem{blakewigglez}
C.~Blake, S.~Brough, M.~Colless, C.~Contreras, W.~Couch, S.~Croom, D.~Croton and T.~Davis {\it et al.},
Mon.\ Not.\ Roy.\ Astron.\ Soc.\ {\bf 425} (2012) 405 B
[arXiv:1204.3674].

 \bibitem{contreraswigglez}
C.~Contreras {\it et al.}  [WiggleZ Collaboration],
Mon.\ Not.\ Roy.\ Astron.\ Soc.\  {\bf 430} (2013) 924C
[arXiv:1302.5178].


\bibitem{beutler6df}
F.~Beutler,
Mon.\ Not.\ Roy.\ Astron.\ Soc.\  {\bf 423} (2012) 3430B
[arXiv:1204.4725].

  
\bibitem{Raccanelli:2012gt}
  A.~Raccanelli, D.~Bertacca, D.~Pietrobon, F.~Schmidt, L.~Samushia, N.~Bartolo, O.~Dore and S.~Matarrese {\it et al.} (2012)
  arXiv:1207.0500.  


\bibitem{Kaiser:1987qv}
  N.~Kaiser,
  Mon.\ Not.\ Roy.\ Astron.\ Soc.\  {\bf 227} (1987) 1.

\bibitem{Szalay:1997cc}
  A.~S.~Szalay, T.~Matsubara and S.~D.~Landy
(1997)  astro-ph/9712007.
  
\bibitem{Matsubara:1999du}
  T.~Matsubara
(1999)  astro-ph/9908056.
  
\bibitem{Szapudi:2004gh}
  I.~Szapudi,
  Astrophys.\ J.\  {\bf 614} (2004) 51
  [astro-ph/0404477].
  
\bibitem{Papai:2008bd}
  P.~Papai and I.~Szapudi (2008)
  arXiv:0802.2940.
  
\bibitem{Raccanelli:2010hk}
  A.~Raccanelli, L.~Samushia and W.~J.~Percival,
   Mon.\ Not.\ Roy.\ Astron.\ Soc.\  {\bf 409} (2010) 1525
  [arXiv:1006.1652].

  
\bibitem{Montanari:2012me}
  F.~Montanari and R.~Durrer,
  Phys.\ Rev.\ D {\bf 86} (2012) 063503
  [arXiv:1206.3545].

\bibitem{Bertacca:2012tp}
  D.~Bertacca, R.~Maartens, A.~Raccanelli and C.~Clarkson,
  JCAP {\bf 1210} (2012) 025
  [arXiv:1205.5221].


\bibitem{Yoo:2009au}
  J.~Yoo, A.~L.~Fitzpatrick and M.~Zaldarriaga,
  Phys.\ Rev.\ D {\bf 80} (2009) 083514
  [arXiv:0907.0707].
  
\bibitem{Yoo:2010ni}
  J.~Yoo,
  Phys.\ Rev.\ D {\bf 82} (2010) 083508
  [arXiv:1009.3021].
 
\bibitem{Bonvin:2011bg}
  C.~Bonvin and R.~Durrer,
  Phys.\ Rev.\ D {\bf 84} (2011) 063505
  [arXiv:1105.5280].
   
\bibitem{Challinor:2011bk}
  A.~Challinor and A.~Lewis,
  Phys.\ Rev.\ D {\bf 84} (2011) 043516
  [arXiv:1105.5292].
  
 

  
\bibitem{Bruni:2011ta}
  M.~Bruni, R.~Crittenden, K.~Koyama, R.~Maartens, C.~Pitrou and D.~Wands,
  Phys.\ Rev.\ D {\bf 85} (2012) 041301
  [arXiv:1106.3999].

\bibitem{Jeong:2011as}
  D.~Jeong, F.~Schmidt and C.~M.~Hirata,
  Phys.\ Rev.\ D {\bf 85} (2012) 023504
  [arXiv:1107.5427].

  
\bibitem{Maartens:2012rh}
  R.~Maartens, G.~-B.~Zhao, D.~Bacon, K.~Koyama and A.~Raccanelli,
  JCAP {\bf 1302} (2013) 044
  [arXiv:1206.0732].
  
 
  \bibitem{Yoo:2012se}
  J.~Yoo, N.~Hamaus, U.~Seljak and M.~Zaldarriaga,
  Phys.\ Rev.\ D {\bf 86} (2012) 063514
  [arXiv:1206.5809].




\bibitem{Hall:2012wd}
  A.~Hall, C.~Bonvin and A.~Challinor,
  Phys.\ Rev.\ D {\bf 87} (2013) 064026
  [arXiv:1212.0728].  
  
\bibitem{Lombriser:2013aj}
  L.~Lombriser, J.~Yoo and K.~Koyama,
  Phys.\ Rev.\ D {\bf 87} (2013) 104019
  [arXiv:1301.3132].
  

\bibitem{Duniya:2013eta}
  D.~Duniya, D.~Bertacca and R.~Maartens,
  JCAP {\bf 1310} (2013) 015
  [arXiv:1305.4509].

\bibitem{Hamilton:1997zq}
  A.~J.~S.~Hamilton
(1997)  astro-ph/9708102.

\bibitem{Hamilton:1995px}
  A.~J.~S.~Hamilton and M.~Culhane,
   Mon.\ Not.\ Roy.\ Astron.\ Soc.\  {\bf 278} (1996) 73
 [astro-ph/9507021].

\bibitem{Bharadwaj:1998bq}
  S.~Bharadwaj,
  Astrophys.\ J.\  {\bf 516} (1999) 507
  [astro-ph/9812274].

 

\bibitem{Raccanelliradial}
A.~Raccanelli, D.~Bertacca, R.~Maartens, C.~Clarkson and O.~Dor\'{e},
(2013) arXiv:1311.6813.


\bibitem{Matarrese:2000iz}
  S.~Matarrese, L.~Verde and R.~Jimenez,
  Astrophys.\ J.\  {\bf 541} (2000) 10
  [astro-ph/0001366].


\bibitem{Dalal:2007cu}
  N.~Dalal, O.~Dor\'{e}, D.~Huterer and A.~Shirokov,
  Phys.\ Rev.\ D {\bf 77} (2008) 123514
  [arXiv:0710.4560].
  
  \bibitem{Matarrese2008}
	S. Matarrese, L. Verde, Astrophys. J. {\bf 677} (2008) L77 [arXiv:0801.4826].

\bibitem{Slosar:2008hx}
  A.~Slosar, C.~Hirata, U.~Seljak, S.~Ho and N.~Padmanabhan,
  JCAP {\bf 0808} (2008) 031
  [arXiv:0805.3580].  


\bibitem{Desjacques:2010jw}
  V.~Desjacques and U.~Seljak,
  Class.\ Quant.\ Grav.\  {\bf 27} (2010) 124011
  [arXiv:1003.5020].
  
  
  \bibitem{Xia:2010pe}
  J.-Q.~Xia, A.~Bonaldi, C.~Baccigalupi, G.~De Zotti, S.~Matarrese, L.~Verde and M.~Viel,
  JCAP {\bf 1008} (2010) 013
  [arXiv:1007.1969].


\bibitem{Matarrese97}  
S.~Matarrese, P.~Coles, F.~Lucchin and L.~Moscardini, 
Mon.\ Not.\ Roy.\ Astron.\ Soc.\  {\bf 286} (1997) 115
[arXiv:astro-ph/9608004].

\bibitem{Moscardini98}  
L.~Moscardini, P.~Coles and F.~Lucchin, Mon.\ Not.\ Roy.\ Astron.\ Soc.\  {\bf 299} (1998) 95
[arXiv:astro-ph/9712184].


\bibitem{Hamaus10}  
N. Hamaus, U. Seljak, V. Desjacques, R. E. Smith and T. Baldauf, 
Phys.\ Rev.\ D {\bf 82} (2010) 043515
[arXiv:1004.5377]

\bibitem{Baldauf11}  
T.~Baldauf, U.~Seljak, L.~ Senatore and M.~ Zaldarriaga,
JCAP {\bf 10} (2011) 031B
[arXiv:1106.5507].

\bibitem{Baldauf13}  
  T.~Baldauf, U.~Seljak, R.~E.~Smith, N.~Hamaus and V.~Desjacques,
  Phys.\ Rev.\ D {\bf 88} (2013) 083507
  [arXiv:1305.2917].




  \bibitem{Xia:2011hj}
  J.-Q.~Xia, C.~Baccigalupi, S.~Matarrese, L.~Verde and M.~Viel,
  JCAP {\bf 1108} (2011) 033
  [arXiv:1104.5015].


\bibitem{Ross:2012sx}
  A.~J.~Ross, W.~J.~Percival, A.~Carnero, G.~-b.~Zhao, M.~Manera, A.~Raccanelli, E.~Aubourg and D.~Bizyaev {\it et al.},
  Mon.\ Not.\ Roy.\ Astron.\ Soc.\  {\bf 428} (2013) 1116
  [arXiv:1208.1491].


\bibitem{Ade:2013ydc} 
  P.~A.~R.~Ade, {\it et al.}  [Planck Collaboration] (2013)
  arXiv:1303.5084.

\bibitem{Camera:2013kpa}
  S.~Camera, M.~G.~Santos, P.~G.~Ferreira and L.~Ferramacho,
  Phys.\ Rev.\ Lett.\  {\bf 111} (2013) 171302
  [arXiv:1305.6928].

\bibitem{Verde:2009hy} 
  L.~Verde and S.~Matarrese,
  Astrophys.\ J.\  {\bf 706}  (2009) L91
  [arXiv:0909.3224].


\bibitem{Linder:2005in}
  E.~V.~Linder,
  Phys.\ Rev.\ D {\bf 72} (2005) 043529
  [astro-ph/0507263].
  
  
\bibitem{Kobayashi:2010wa}
  T.~Kobayashi,
  Phys.\ Rev.\ D {\bf 81} (2010) 103533
  [arXiv:1003.3281].

  
  

\end{thebibliography}
\end{document}